\documentclass[11pt]{article}
\usepackage{amsmath,amssymb,bm,epsf,epsfig,graphicx}
\input epsf.sty
\usepackage[utf8]{inputenc}
\usepackage[english]{babel}
\usepackage{amsmath}
\usepackage{latexsym}
\usepackage{amsfonts}
\usepackage{mathtools}
\usepackage{amssymb}
\usepackage{scalerel}
\usepackage{extpfeil}
\usepackage[left=3cm,right=3cm,top=3.1cm,bottom=3.1cm]{geometry}
\usepackage{cite}
\usepackage{tensor}
\usepackage{tikz}
\usepackage{tikz-cd}
\usepackage{bbm}

\usepackage[mathscr]{euscript}
\usetikzlibrary{arrows.meta, bending}
\tikzset{C/.style={circle, minimum size=8mm,
		node contents={},
		append after command={\pgfextra{%
				\draw[-{Straight Barb[flex']}](\tikzlastnode.150) arc (450:110:2.8mm);}
	}}
}

\usepackage{hyperref}
\numberwithin{equation}{section}

    \newcommand{\Tr}{\mathop{\rm Tr}\nolimits}

    \def\bra#1{\langle #1 |}
    \def\ket#1{|#1 \rangle}

    \def\p{\partial}

    \def \be {\begin{eqnarray}}
    \def \ee {\end{eqnarray}}
    \def \bal {\begin{align}}
    \def \eal {\end{align}}
    \def \bdm {\begin{displaymath}}
    \def \edm {\end{displaymath}}

    \def\del {\partial}
    \def\0{\nonumber}

\begin{document}
	\begingroup\allowdisplaybreaks

\vspace*{1.1cm}
\centerline{\Large \bf Boundary Modes in String Field Theory}


\vspace{.3cm}

\begin{center}

{\large Carlo Maccaferri$^{(a)}$\footnote{Email: maccafer at gmail.com}, Riccardo Poletti$^{(a,b)}$\footnote{Email: polettiricc at gmail.com }, Alberto Ruffino$^{(a)}$\footnote{Email: ruffinoalb at gmail.com}  and  Jakub Vo\v{s}mera$^{(c)}$\footnote{Email: jakub.vosmera at ipht.fr} }
\vskip 1 cm
$^{(a)}${\it Dipartimento di Fisica, Universit\`a di Torino, \\INFN  Sezione di Torino \\
Via Pietro Giuria 1, I-10125 Torino, Italy}
\vskip .5 cm
\vskip .5 cm
$^{(b)}${\it CEICO, Institute of Physics, Czech Academy of Sciences\\
Na Slovance 1999/2, Prague, 182\ 00, Czechia}
\vskip .5 cm
\vskip .5 cm
$^{(c)}${\it Institut de Physique Théorique\\
	CNRS, CEA, Université Paris-Saclay\\
 Orme des Merisiers, Gif-sur-Yvette, 91191 CEDEX, France}

%
\end{center}

\vspace*{6.0ex}

\centerline{\bf Abstract}
\bigskip
We discuss the construction of boundary contributions to free string field theory actions in the context of the bosonic string. We show that it is generally possible to obtain a well-defined variational principle by adding a simple boundary term (which only depends on the value of the bulk fields at the boundary) to the original bulk action. However, it is in general not possible to do this in a gauge-invariant way unless suitable boundary degrees of freedom are added. We explicitly construct such boundary contributions for the massless level of both the open and the closed SFT, as well as for the tensionless limit of the full string field theory. Our results reproduce linearized general relativity with the Gibbons-Hawking-York term and provide similar gauge-invariant actions for the infinite tower of massless higher-spin gauge theories for all Regge trajectories. By writing down a gauge-invariant action for the first massive level of the open string, we provide evidence that an analogous construction should be possible for the full tensile string field theory. 

\baselineskip=16pt
\newpage
\setcounter{tocdepth}{2}
\tableofcontents

\section{Introduction}\label{sec:1}

There are  indications that  the current formulation of string field theory (SFT, see \cite{Sen:2024nfd, Maccaferri:2023vns}  for recent reviews  and references) is incomplete.  In \cite{Erler:2022agw}, for example, it was shown that classical solutions in closed SFT have a vanishing action, at least in compact target spaces. 
Moreover, in a possible attempt to describe black holes from SFT (perhaps following the pioneering approach of \cite{Mukherji:1991kz} or by some other means) one is naturally lead to asking what kind of SFT gauge-invariant quantity would compute the ADM observables of the black-hole. It seems natural to expect that these fundamental questions can be addressed by adding boundary contributions to the SFT action. 
After all, string theory is a theory of gravity and the standard Einstein-Hilbert (EH) action needs the Gibbons-Hawking-York (GHY) boundary term \cite{York:1972sj, Gibbons:1976ue} to properly account for measurable quantities, like the ADM observables\cite{Arnowitt:1962hi}  which, for black hole solutions, are fundamentally captured by the GHY term. This paper contains some preliminary non-trivial results in this direction. 

The covariant free action for string field theory on a flat $d$-dimensional target space reads\footnote{See the beginning of section \ref{sec:position} for detailed description of the string field $\Psi(x)$ in position space and the reduced symplectic form $\omega'$.}
\begin{align}
\frac12\omega(\Psi,\hat Q\Psi)=\frac12\int d^d x\,\omega'\!\left(\Psi(x),\,Q\Psi(x)\right),\label{psiQpsi}
\end{align}
where $Q\sim -c\partial_\mu\partial^\mu +\cdots$ is the kinetic operator whose leading contribution contains a double derivative. To get the standard equation of motion $\hat Q\Psi=0$, one needs to integrate by parts the $\partial_\mu$ derivatives. This produces boundary terms which are usually ignored, either assuming that the string fields $\Psi(x)$ vanish fast enough at infinity or simply considering a compact target space with no boundaries. If we try to define the action  \eqref{psiQpsi} on a manifold $M$ with boundary $\partial M$ (by simply replacing $\int\to\int_M$) we get problems, first of all, with the variational principle. 
 To appreciate this, consider a simple (euclidean) massless scalar field theory analogous to \eqref{psiQpsi}
\begin{align}
S(\phi)=\frac12\int_M d^dx \,\phi(-\partial_\mu\partial^\mu)\phi\,.
\end{align}
Integrating by parts and using the divergence theorem 
$$
\int_M d^d x\,\partial_\mu f^\mu(x)=\int_{\partial M} d^{d-1}y\sqrt{\gamma}\, n_\mu(y) f^\mu(x(y))\,,
$$
(where $x(y)$ describe the embedding of $\partial M$ into $M\subset \mathbb R^d$, $\gamma$ is the determinant of the induced metric and $n^\mu(y)$ is the normal unit vector of $\partial M$, pointing outside $M$)
the variation of the action gives
\begin{align}
\delta S(\phi)=&\,-\frac12\int_M d^dx \,\left(\delta\phi(-\partial_\mu\partial^\mu)\phi + \phi(-\partial_\mu\partial^\mu)\delta\phi\right)\nonumber\\
=&\,\int_M d^dx\,\delta\phi\left(-\partial_\mu\partial^\mu\phi\right)+\0\\
&\,-\frac{1}{2}\int_{\partial M} d^{d-1}y \,\sqrt{\gamma}\,n_\mu\,\delta\phi\,\partial^\mu\phi+\frac{1}{2}\int_{\partial M} d^{d-1}y \,\sqrt{\gamma}\,n_\mu\phi\,\delta\partial^\mu\phi\,.\label{varphi}
\end{align}
The bulk part of the variation gives the standard wave equation $\del^2\phi=0$. The first boundary contribution to the variation can be set to zero by the Dirichlet 
$$\delta \phi{\Big |}_{\partial M}=0\,,
$$ 
or the Neumann boundary conditions
$$
n_\mu\del^\mu \phi{\Big |}_{\partial M}=0\,,
$$  
which are consistent boundary conditions for the wave equation. After this, however, it is not possible to set to zero the second boundary contribution in \eqref{varphi} without over-constraining the space of solutions. Thus the variational principle is ill-defined. If, on the other hand, we started with the naively equivalent action with only first derivatives  
\begin{align}
S'(\phi)=\frac12\int_M d^dx \,\partial_\mu\phi\,\partial^\mu\phi\,,
\end{align}
we would have found 
\begin{align}
\delta S'(\phi)=\,\int_M d^dx\,\delta\phi\left(-\partial_\mu\partial^\mu\phi\right)
-\int_{\partial M} d^{d-1}y \,\sqrt{\gamma}\,n_\mu\,\delta\phi\,\partial^\mu\phi\,,
\end{align}
which, according to what we have just reviewed, constitutes a well-defined variational principle.

Similarly, the gauge invariance of the action \eqref{psiQpsi} under the standard SFT gauge transformation $\delta\Psi=\hat Q\Lambda$ requires that $\hat Q$ is cyclic with respect to the symplectic form $\omega$, which in practice means that it should be possible to integrate $\partial_\mu$ by parts without producing boundary terms. This is not possible in presence of a boundary.

 To summarize, if defined on a target space with boundary, the action \eqref{psiQpsi} has two serious deficiencies:
\begin{enumerate}
\item it does not give a well-defined variational principle,
\item it is not gauge-invariant.
\end{enumerate}
It turns out that it is rather easy to modify the action \eqref{psiQpsi} so that it gives a well-defined variational principle. Essentially this amounts to adding a suitable boundary term which would facilitate integrating by parts one $\del_\mu$ from the kinetic operator $\del_\mu\del^\mu$. In the above example of the massless scalar, this would correspond to adding to $S(\phi)$ a boundary term so that it coincides with $S'(\phi)$, namely
\begin{align}
S(\phi)\to S'(\phi)= S(\phi)+\frac12 \int_{\partial M} d^{d-1}y \sqrt{\gamma}n_\mu\left(\phi\partial^\mu\phi\right)\,.
\end{align}
As we will see, an analogous boundary contribution can be added in a multitude of ways to the naive SFT action \eqref{psiQpsi} in order to make the variational principle well-defined.

However, the absence of gauge invariance of the action \eqref{psiQpsi} turns out to be a more delicate issue: a simple recipe, like the one just described, will not generally be capable of restoring it. Indeed, the failure of gauge invariance is fundamentally not related to the fact of having second order derivatives: it is simply a consequence of having a boundary. To get a first glimpse of this problem (and, at the same time, a crucial hint for how to solve it),  let us consider the abelian Chern-Simons theory on a 3-manifold $M$
\begin{align}
S_\text{CS}(A)=\frac12\int_M d^3x \,\epsilon^{\mu\nu\rho}\,A_\mu\del_\nu A_\rho=\frac12\int_{M}\,A\wedge\,dA\,.\label{eq:SCS}
\end{align}
The action contains only first derivatives so it gives a well-defined variational principle. Furthermore, neglecting boundary contributions, one would have expected the action should be left invariant under the obvious gauge transformation 
\begin{align}
\delta_\xi A=d\xi\,.\label{eq:gauge_CS}
\end{align}
However, carefully keeping track of all boundary terms which are picked up when integrating by parts, one finds the gauge variation of the action is, in fact, not vanishing 
\begin{align}
\delta_\xi S_\mathrm{CS}(A)=&\,\frac12\int_M  d\xi\wedge dA=\frac12\int_M d\left(\xi\wedge  A\right)=\frac12\int_{\del M}\,\xi\wedge dA\neq0\,.
\end{align}
Since gauge transformations are just redundancies in language, they cannot be broken by adding a boundary. In other words, if we are to consistently add a boundary to a gauge theory, there should be a way of doing it in a gauge-invariant manner. 
In the above case of the abelian Chern-Simons theory, a solution is immediately visible and it is given by introducing a boundary degree of freedom which, when endowed with an appropriate gauge transformation law, will compensate for the non-invariance of the bulk action. This then gives rise to physical excitations propagating on the boundary, namely the {\it edge modes} \cite{Witten:1988hf,Wen:1992vi} (see \cite{Tong:2016kpv} for a review).
In particular, the minimal boundary term which is capable of restoring gauge-invariance of the action, turns out to involve an extra scalar field $m$ living on the boundary, namely
\begin{align}
S_{\p M}(A,m)=-\frac12\int_{\del M} \, m\wedge dA\,,
\end{align}
where we equip $m$ with the gauge transformation
\begin{align}
\delta_\xi m=\xi\,.\label{eq:gauge_m}
\end{align}
A fully gauge-invariant action for the abelian Chern-Simons theory with boundary can thus be written as
\begin{align}
S'_\mathrm{CS}(A,m)
=S_\mathrm{CS}(A)+S_{\p M}(A,m)=\frac12\int_M\, A\wedge dA-\frac12\int_{\p M}\,m\wedge dA\,.\label{eq:SCS_gauge_inv}
\end{align}
The original action \eqref{eq:SCS} can then be recovered from \eqref{eq:SCS_gauge_inv} by gauging the field $m$ away. Doing so, we completely fix the gauge parameter at the boundary (as can be seen from \eqref{eq:gauge_m}), meaning that \eqref{eq:SCS} can be regarded as a partially gauge-fixed action, which is invariant under \eqref{eq:gauge_CS} provided that $\xi$ is restricted to vanish at $\p M$. From the perspective of the action \eqref{eq:SCS}, the boundary degrees of freedom are then hidden inside the field configurations of the gauge field $A$ which would have been gauge-equivalent in the bulk, but not so on the boundary. 

It is of particular interest to isolate the dynamics of the physical boundary excitations and investigate their relation to the compensating boundary field $m$ appearing in the gauge-invariant action \eqref{eq:SCS_gauge_inv}. For the sake of concreteness, let us take $M$ to be a 2+1 dimensional flat half-space with coordinates $(t,y,z)$, which is bounded by a flat space-like codimension-1 hyperplane $\p M$ sitting at $z=0$. The bulk of $M$ extends for $z<0$. In this parametrization, the action \eqref{eq:SCS_gauge_inv} explicitly reads
\begin{align}
    S'_\mathrm{CS}(A,m)= \frac{1}{2}\int_M d^3 x\, \epsilon^{ijk} A_i \p_j A_k -\frac{1}{2}\int_{\p M} dt\,dy\, (\p_t A_y - \p_y A_t)\, m\,,\label{eq:Styz}
\end{align}
where $i,j,\ldots\in \{t,y,z\}$. Let us then consider integrating out the time-like component $A_t$ of the gauge field. Varying the action \eqref{eq:Styz} with respect to $A_t$, one obtains a bulk term which imposes the equation of motion
\begin{align}
    \p_y A_z - \p_z A_y=0\,,\label{eq:eomAt}
\end{align}
as well as a boundary term, which can be set to zero by requiring $A_t$ to satisfy the (gauge-invariant) Dirichlet boundary condition
\begin{align}
    A_t = \p_t m\label{eq:At_bc}
\end{align}
at $z=0$. The equation of motion \eqref{eq:eomAt} can generally be solved by setting
\begin{subequations}
\label{eq:Axy_on_shell}
\begin{align}
    A_y &= \p_y f\,,\\
    A_z &= \p_z f\,,
\end{align}
\end{subequations}
for some scalar field $f$ which simply shifts as $\delta_\xi f = \xi$ under a gauge transformation. Substituting the on-shell values \eqref{eq:Axy_on_shell} and \eqref{eq:At_bc} of the gauge field $A$ back into the action \eqref{eq:Styz} and realizing that the coordinates $t$ and $y$ can be integrated by parts with impunity, one obtains
\begin{subequations}
    \begin{align}
        S'_\mathrm{CS}(A^\ast,m) &=\frac{1}{2}\int_M d^3 x\, (\p_z A_t-\p_t \p_z f)\,\p_y f +\frac{1}{2}\int_M d^3 x\,   (\p_t\p_y f-\p_y A_t)\,\p_z f+\nonumber\\
        &\hspace{6cm}+\frac{1}{2}\int_{\p M} dt\,dy\, m \,\p_t\p_y( m - f)\\
        &=\frac{1}{2}\int_M d^3 x\, \p_z( A_t \p_y f+f\p_t\p_y f) +\frac{1}{2}\int_{\p M} dt\,dy\, m \,\p_t\p_y( m - f)\\
         &=\frac{1}{2}\int_{\p M} dt\,dy\, (m-f)\,\p_t\p_y (m-f)\,,
    \end{align}
\end{subequations}
where we have realized that $A_t$ ends up contributing only through boundary terms so that in the end, it can be always replaced by its boundary on-shell value \eqref{eq:At_bc}.
Hence, introducing the gauge-invariant scalar edge mode
\begin{align}
    \phi\coloneqq m-f\,,
\end{align}
the abelian Chern-Simons action \eqref{eq:SCS_gauge_inv} reduces to the Floreanini-Jackiw action\footnote{In the literature, one usually encounters the form
\begin{align}
    \frac{1}{2}\int_{\p M} dt\,dy\, \phi\,(\p_t\p_y -v\p_y^2)\phi
\end{align}
of the Floreanini-Jackiw action. This can be obtained from \eqref{eq:FJ} by the simple Galilean boost $t\to t$, $y\to y+vt$.}
\begin{align}
   S_\mathrm{edge}(\phi) =\frac{1}{2}\int_{\p M} dt\,dy\, \phi\,\p_t\p_y \phi\label{eq:FJ}
\end{align}
describing a chiral boson in the (1+1)-dimensional space $\p M$.
 
Another important classical example of a boundary term that makes the variational principle well-defined and, at the same time, restores gauge invariance, is the Gibbons-Hawking-York (GHY) term of General Relativity \cite{York:1972sj, Gibbons:1976ue}
\begin{align}
    S_\mathrm{GR}= S_\mathrm{EH}+S_\mathrm{GHY}= \int_M d^d x\,\sqrt{g}\, R +\int_{\p M} d^{d-1} y\,\sqrt{\gamma}\, 2K,\label{GHY}
 \end{align}
where $K$ is the extrinsic scalar curvature of the boundary. As we will review in section \ref{subsec:comments}, this boundary term not only makes the variational principle for massless spin-2 fluctuations well-defined: by considering variations of $\p M$ (i.e.\ a boundary degree of freedom) in addition to fluctuations of the bulk metric, full invariance under infinitesimal diffeomorphism gauge transformations can be restored. This, for instance, can be explicitly seen by expanding the full action \eqref{GHY} around a flat boundary in flat space, as will be done in appendix \ref{app:linear}, where a specific boundary field $l(y)$ \eqref{eq:fl} associated with the transverse displacement of the boundary will be identified. Without this degree of freedom, which gauge-transforms with the component of the bulk diffeomorphism which is normal to $\p M$, the action would not have been invariant under generic diffeomorphisms: fixing the boundary to be rigid (i.e.\ gauging away $l$) would result in a partially gauge-fixed action, with the residual gauge symmetry being given by diffeomorphisms which respect the boundary. In other words, full diffeomorphism invariance is only realized if the boundary is left free to fluctuate, with the field $l$ playing an analogous role as the field $m$ in the Chern-Simons action \eqref{eq:SCS_gauge_inv}.

Since gravity generally constitutes a subsector of string theory, we should expect that analogous boundary degrees of freedom should be needed to consistently define SFT on a target space with boundary.\footnote{In general, note that if we have the luxury of  a bulk action  containing only first derivatives and  built using gauge-invariant (or covariant) blocks (such as the integrand $\Tr [F_{\mu\nu}F^{\mu\nu}]$ of the Yang-Mills action) then boundary degrees of freedom are not needed. However, except for the photon/gluon and the Kalb-Ramond 2-form, this does not generically happen in (bosonic) string theory.} The introduction of such boundary modes, for the time being restricted to the massless sector and the tensionless limit of the full string theory, is the main contribution of this paper.

Let us summarize our main results by giving the plan of the paper. In section \ref{sec:position} we first of all set out our conventions and we discuss the position representation of string fields. This turns out to be a rather useful language in the cases when the matter CFT representing the string background contains $d$ non-compact free-bosons $X^\mu(z,\bar z)$, with $\mu=1,\ldots, d$. Then we discuss how the symplectic form used to construct the kinetic term \eqref{psiQpsi} is modified by the addition of a boundary and how this implies that the BRST operator $Q$ fails to be cyclic. We show how this is related to the fact that the variational principle is ill-defined and we give an infinite family of boundary terms which rectifies this issue. This redundancy boils down to the choice of an operator $\Gamma$ and its bpz conjugate $\Gamma^*$ (see \eqref{eq:GammaDecomp}) which, together, measure the failure of the BRST operator to be cyclic.  
We then discuss the possibility of choosing a boundary term from the discovered infinite family so as to maintain the original gauge-invariance, ending up with an operator equation $\Gamma^* Q=0$ which does not generally appear to have non-trivial solutions. We conclude the section by drawing a gravity-inspired general consideration on the need of introducing additional boundary degrees of freedom. 
In section \ref{sec:massless}, we systematically build a consistent free action for the massless sector of the bosonic string field theory in a flat bulk space with a flat boundary. This should be an example of a simple starting background which is free from boundary tadpoles from the gravity perspective. Thanks to the addition of appropriate boundary string fields subject to an algebraic (gauge-invariant) constraint (as discussed in section \eqref{subsec:boundary_sf}), we will finally write down a bulk+boundary gauge-invariant action with well-defined variational principle for both the free open, as well as free closed strings at the massless level. Although our algebraic setting fails short of applying to the full massive string theory, we show in section \ref{sec:tensionless} that it can easily accommodate the tensionless limit of both open and closed bosonic string. We therefore write down, for the first time to our knowledge, the needed boundary contributions to render the Fronsdal description \cite{Fronsdal:1978rb} of massless higher spins well-defined and gauge invariant in the presence of a (flat) boundary.  Finally, in section \ref{sec:disc}, we discuss the extension of our results to the full massive sector, concluding that a non-trivial extension of our present construction is needed. Then, anticipating results from a forthcoming work \cite{future}, we write down a gauge invariant free action for the states of the first massive level of the open string and for the corresponding boundary modes. Two appendices contain important complementary material. In appendix \ref{app:linear} we linearize the combined Einstein-Hilbert and Gibbons-Hawking-York action around flat space with a flat boundary. Doing this, we make contact with the massless sector of closed string theory, as well as with the spin-2 sector of the tensionless open string theory, finding perfect agreement in the spectrum and action of both the bulk and the boundary degrees of freedom. In Appendix \ref{app:nilpotent} we reformulate our gauge-invariant action in terms of a single bulk-boundary nilpotent structure, which should be a good starting point for the construction of the interactions and possibly the quantum extension.

 {\bf Note added}: While this work was in progress, the papers \cite{georg, atakan} appeared which discuss some of these problems from the viewpoint of the worldsheet path-integral \cite{Kraus}  and partially overlap with our results up to Section \ref{subsec:restoring_cyc}. 

\section{Position-space approach to SFT}\label{sec:position}
In this section, we introduce a general formalism applicable in the case of a free string field theory on a flat non-compact $d$-dimensional target space. Adding a codimension-one boundary, we will demonstrate the failure of the naive SFT kinetic term to provide a well-defined gauge-invariant action for free target fields. Throughout the paper, in order to avoid subtleties associated with time-like and null boundaries, we will always have in mind the euclidean signature $\eta_{\mu\nu}=\mathrm{diag}[+,\ldots,+]$ for all non-compact directions.

\subsection{Notation and conventions}

Let us first set up the scene by outlining a target-oriented notation which is to be used throughout the paper. 

\subsubsection{String fields in position space}

A generic string field in the aforementioned setup can be expanded in the basis of the momentum eigenstates as
\begin{align}
    |\Psi\rangle = \sum_s\int_{\mathbb{R}^d} \frac{d^d k}{(2\pi)^d} \tilde{\Psi}_s(k)\alpha_s |k\rangle\,,\label{eq:SF}
\end{align}
where the operators $\alpha_s$ collectively denote matter and ghost oscillators such that $\alpha_s\ket k\neq0$. Note that the matter sector may in general include an internal (compact) sector which is tensored with the CFT of the $d$ non-compact free bosons so as to saturate the ghost central charge $c_\text{gh}=-26$.
The momentum eigenstates $|k\rangle$ satisfy
\begin{align}
    e^{ik\cdot X(0)}|0\rangle =|k\rangle = e^{ik\cdot \hat{x}}|0\rangle\,,
\end{align}
where $|0\rangle$ is the zero-momentum vacuum ($SL(2,\mathbb{R})$ in the case of the open string, $SL(2,\mathbb{C})$ for the closed string) and $\hat{x}$ denotes the position zero mode. 
This satisfies the canonical commutation relation
\begin{align}
[\hat{x}^\mu,\hat{p}^\nu]=i\eta^{\mu\nu}\,,
\end{align}
with the corresponding conjugate momentum operator $\hat{p}^\mu$. Hence, denoting by $\Psi_s(x)$ the Fourier transform
\begin{align}
\Psi_s(x)     = \int_{\mathbb{R}^d} \frac{d^d k}{(2\pi)^d} \, \tilde{\Psi}_s(k) \, e^{ik\cdot {x}}
\end{align}
of the momentum-space expansion coefficients $\tilde{\Psi}_s(k)$, the string field \eqref{eq:SF} can be recast as a function of the zero modes $\hat{x}$ as
\begin{align}
    |\Psi\rangle = \Psi(\hat{x})|0\rangle\,,
\end{align}
where 
\begin{align}
    \Psi(\hat{x}) = \sum_s \Psi_s(\hat{x}) \,\alpha_s\,.\label{eq:oscExp}
\end{align}

\subsubsection{BPZ inner product and symplectic form}

On this space of string fields, let us consider the BPZ inner product
\begin{align}
\langle\Phi |\Psi\rangle=\langle 0| \Phi^*(\hat{x})\Psi(\hat{x}) |0\rangle\,,
\end{align}
where  $^*$ denotes the BPZ conjugation. This product can be evaluated in terms of the target fields $\Psi_s(x)$  
by inserting the completeness relation
$$
\int_{\mathbb{R}^d} d^dx \,\ket x \bra x =\mathbb{I}
$$
and noting that we can write
$$
\bra x 0\rangle=\lim_{k\to 0} \bra x k\rangle=\lim_{k\to 0} e^{i k\cdot x}=1\,.
 $$
One then obtains
\begin{align}
   \langle\Phi |\Psi\rangle =\sum_{s,s'}  \int_{\mathbb{R}^d}d^dx\, W_{ss'}\,\Phi_s(x) \Psi_{s'}(x)\,,
\end{align}
with the matrix elements $ W_{ss'}\equiv \langle 0|\alpha_s^* \alpha_{s'}| 0\rangle$.
Appropriately assigning a degree $d(\Psi)\in \mathbb Z$ to the generic string field $\ket\Psi$ based on its ghost number, we can furthermore introduce the graded-antisymmetric product (a.k.a.\ symplectic form)
\begin{align}
\omega(\Phi,\Psi)=(-1)^{d_\Phi}\bra\Phi\# \ket\Psi
\end{align}
where $\#$ is a placeholder for a possibly-needed ghost insertion: nothing in the case of OSFT and $c_0^-\equiv \frac{1}{2}( c_0-\bar{c}_0)$ for CSFT. 

\subsubsection{Primed symplectic form}

In the following, we will find it very convenient to make explicit the target integration over $x$ by writing
\begin{align}
    \omega(\Phi,\Psi)=\int_{\mathbb{R}^d}d^d x\, \omega'(\Phi(x),\Psi(x))\,,
\end{align}
where $\omega'$ is the obvious reduced symplectic form for the states in which the position zero-mode operators $\hat{x}$ were replaced by their eigenvalues $x$ 
\begin{align}
    \Psi(\hat{x}) \longrightarrow \Psi(x)\,.
\end{align}
More explicitly, given that we decompose
$$
 \Psi({x}) = \sum_s \Psi_s({x}) \,\alpha_s\,,
$$
(similarly as in \eqref{eq:oscExp} but now with all fields evaluated at $x$, not at the operator $\hat{x}$), we can define the primed symplectic form as
$$
\omega'(\Phi(x),\Psi(x)) \equiv (-1)^{d_\Phi} \langle 0|\Phi^*(x)\# \Psi(x)|0\rangle = (-1)^{d_\Phi} \sum_{s,s'} \omega_{ss'} \,\Phi_s(x) \Psi_{s'}(x) \,.
$$
with the matrix elements $ \omega_{ss'}\equiv \langle 0|\alpha_s^* \# \alpha_{s'}| 0\rangle$.

\subsection{Case with no boundaries}

Let us start by writing down the free action in the flat space $\mathbb{R}^d$.

We will be assuming suitable fall-off conditions for the target fields $\Psi_s(x)$ so that one can integrate by parts without picking up any boundary contributions at infinity. 
In particular, for any field configurations $\Psi_s(x),\Psi_{s'}(x)$ (as well as for all their derivatives), we will assume that
\begin{align}
\int_{\mathbb{R}^d} d^dx \, \Psi_s(x)\,\partial_\mu\Psi_{s'}(x)=-\int_{\mathbb{R}^d} d^dx \, (\partial_\mu\Psi_s(x))\Psi_{s'}(x)\,. \label{eq:per_partes}
\end{align}
Eventually, we will be interested in defining a consistent field theory on a domain $M\subset\mathbb{R}^d$, where we would like to allow the fields $\Psi_s(x)$ to take generic non-zero values on $\p M$. To this end, we will assume that the fall-off conditions which guarantee the validity of \eqref{eq:per_partes} do not generally limit possible field configurations in the parts of $M$ and its boundary $\partial M$ which stay away from infinity.

The free action for $\Psi_s(x)$ then reads
\begin{align}
S_0(\Psi)=\frac{1}{2}\omega(\Psi,\hat{Q}_\mathrm{B}\Psi)=\frac{1}{2}\int_{\mathbb{R}^d} d^dx \, \omega'(\Psi(x),Q_\mathrm{B}\Psi(x))\,,\label{eq:actionCompact}
\end{align}
where $\hat{Q}_\mathrm{B}$ denotes the nilpotent BRST operator. 
Inside the primed symplectic form, the BRST charge can be thought of as acting in the position representation (denoted by $Q_\mathrm{B}$, without the hat) and will be assumed to take the generic form
\begin{align}
Q_\mathrm{B}=-\alpha' c\,\partial_\mu\partial^\mu- \Omega^\mu \partial_\mu+Q'\,.\label{eq:Qdecomp}
\end{align}
The decomposition \eqref{eq:Qdecomp} is defined so that $\Omega^\mu$ and $Q'$ contain neither $x^\mu$ nor the target derivatives $\p_\mu$. In particular, $c$ denotes a $c$-ghost zero-mode\footnote{$c_0$ for OSFT, $\frac{1}{2}c_0^+ = \frac{1}{4}(c_0+\bar{c}_0)$ for CSFT}, while $\Omega^\mu$ and $Q'$ are certain linear combinations of matter and ghost oscillators with ghost-number one, for which the nilpotency of $Q_\mathrm{B}$ implies the algebra
\begin{subequations}
\label{eq:Qalgebra}
\begin{align}
    c^2&=0\,,\\
    [c,\Omega^\mu]&=0\,,\\
    -\alpha'\eta^{\mu\nu}[c,Q']+\tfrac{1}{2}[\Omega^\mu,\Omega^\nu]&=0\,,\\
    [Q',\Omega^\mu]&=0\,,\\
    (Q')^2&=0\,.
\end{align}
\end{subequations}
Furthermore, we have the BPZ properties $c^*=-c$, $\Omega_\mu^\ast=+\Omega_\mu$, as well as $(Q')^\ast=-Q'$.
As a consequence of these, the BRST charge \eqref{eq:Qdecomp} is a BPZ-odd operator, that is 
\begin{align}
    Q_\mathrm{B}^*=-Q_\mathrm{B}\,.\label{eq:Qodd}
\end{align}
However, since $Q_\mathrm{B}$ contains derivatives with respect to target coordinates, the \emph{cyclicity} of $\hat{Q}_\mathrm{B}$ with respect to $\omega$ crucially depends on the ability to integrate by parts without getting contributions at infinity. Indeed, it is only after making use of the fall-off conditions \eqref{eq:per_partes} of $\Psi_s(x)$ at infinity that the property \eqref{eq:Qodd} guarantees that $\hat{Q}_\mathrm{B}$ is cyclic with respect to $\omega$, namely
\begin{align}
\omega(\Psi_1,\hat{Q}_\mathrm{B}\Psi_2)=-(-1)^{d({\Psi_1})}\omega(\hat{Q}_\mathrm{B}\Psi_1,\Psi_2)\,.\label{cycl}
\end{align} 
A generic variation of the action \eqref{eq:actionCompact} is therefore given by
\begin{align}
\delta S_0=\omega(\delta\Psi,\hat{Q}_\mathrm{B}\Psi)=\int_{\mathbb{R}^d}d^d x\,\omega'(\delta\Psi(x),Q_\mathrm{B}\Psi(x))\,,
\end{align}
with no extra boundary terms. This implies both the well-definedness of the variational principle given by the action \eqref{eq:actionCompact}, as well as its invariance under the gauge transformation $\delta_{\Lambda}\Psi=\hat{Q}_\mathrm{B}\Lambda$.

\subsection{Introducing a boundary}

We would now like to write down a free action for the target fields $\Psi_s(x)$ on a domain $M\subset \mathbb{R}^d$ with a codimension-one boundary $\p M$ where the fields $\Psi_s(x)$ are not constrained by any fall-off conditions at $\p M$. In particular, we will no longer assume that integration by parts can be performed inside $M$ without picking up contributions localized at $\p M$. On the other hand, for the sake of simplicity, we will focus on configurations, where integrating by parts over $\p M$ produces no additional (corner) terms. This can be ensured either by assuming that $\p(\p M)=\emptyset$ or by adopting suitable fall-off conditions towards the boundaries of $\p M$ (corners of $M$). 

\subsubsection{Boundary operator}

A naive candidate for such an action would be obtained by replacing
$$
\omega \longrightarrow \omega_M\,,
$$
in the action \eqref{eq:actionCompact}, where the symplectic form $\omega_M$ can be defined in terms of $\omega$ with an extra insertion of the indicator function $\theta$ pointing on the inside of $M$, evaluated at $\hat{x}$. In particular, the distribution $\theta(x)$ is defined to restrict integrals over $\mathbb{R}^d$ to those over $M$, namely
\begin{align}
    \int_{\mathbb{R}^d} d^d x \,\theta(x)\, f(x) = \int_{M} d^d x \, f(x)
\end{align}
for any function $f$ on $\mathbb{R}^d$.
The symplectic form $\omega_M$ can then be expressed as
\begin{align}
    \omega_M(\Phi,\Psi) \equiv \omega(\Phi,\theta(\hat{x})\Psi)= \int_{M} d^dx \, \omega'(\Phi(x),\Psi(x))\,.
\end{align}
The failure of the ability to integrate by parts on $M$ without obtaining extra boundary terms is now reflected in $\hat{Q}_\mathrm{B}$ being no-longer cyclic with respect to $\omega_M$. Equivalently, one can say that the operator $\theta(\hat{x})\hat{Q}_\mathrm{B}$ is not BPZ-odd because $(\theta(\hat{x})\hat{Q}_\mathrm{B})^\ast = -\hat{Q}_\mathrm{B}\theta(\hat{x})\neq -\theta(\hat{x})\hat{Q}_\mathrm{B}$ and therefore not cyclic with respect to $\omega$. This failure of cyclicity can be parametrized by the relation 
\begin{align}
\omega_M(\Psi_1,\hat{Q}_\mathrm{B}\Psi_2)=-(-1)^{d({\Psi_1})}\omega_M(\hat{Q}_\mathrm{B}\Psi_1,\Psi_2)+\omega(\Psi_1,\hat{B}\Psi_2)\,,
\end{align} 
where we have introduced the \emph{boundary operator}\footnote{A similar expression (replacing $\hat{x}$ with the full oscillator expansion of $X(z,\bar{z})$ inserted at the open-string midpoint) appeared in \cite{Witten:1986qs} and \cite{Cho:2023khj} in the context of the construction of a symplectic structure of OSFT. See also \cite{Balasubramanian:2018axm}.}
\begin{align}
    \hat{B}\equiv [\theta(\hat{x}),\hat{Q}_\mathrm{B}]\,.\label{eq:Bdef}
\end{align}
Note that such $\hat{B}$ is BPZ even, namely $\hat{B}^\ast=\hat{B}$, and cyclic with respect to $\omega$. Moreover, when acting on the oscillator Hilbert space inside the primed symplectic form, $B$ can be expanded as
\begin{align}
    B\,\cdot  = \alpha'c\big[\delta_\mu(x)\p^\mu\!\cdot +\p^\mu(\delta_\mu(x)\,\cdot)\big] + \Omega^\mu \delta_\mu \,\cdot \,,
\end{align}
where $\cdot$ is a placeholder for the state on which $B$ acts. Here we have defined
\begin{align}
    \delta_\mu(x) = \p_\mu \theta(x) = -n_\mu(y)\,\delta(x)\,,
\end{align}
where $y^a$ denote coordinates on the manifold $\p M$, which is now understood as being described by the embedding $x^\mu=x^\mu(y^a)$. 
Moreover, $n^\mu(y)$ is the outward-pointing unit vector normal to $\p M$ and the distribution $\delta(x)$ localizes on $\p M$ in the sense that
\begin{align}
    \int_{\mathbb{R}^d} d^d x \,\delta(x)\, f(x) = \int_{\p M} d^{d-1} y\, \sqrt{\gamma}\, f(x(y))\,,
\end{align}
for any function $f$ on $\mathbb{R}^d$ where $\gamma_{ab}=\partial_a x^\mu(y)\,\partial_b x^\nu(y) \,\eta_{\mu\nu}$ is the induced metric on $\p M$. 


\subsubsection{Naive action}\label{naive action}

It follows that if we were to consider the action of the form
\begin{align}
S_0 (\Psi)=\frac12\omega_M(\Psi,\hat{Q}_\mathrm{B}\Psi)=\int_{M}d^d x\, \omega'(\Psi(x),Q_\mathrm{B}\Psi(x))\,,\label{eq:actionNaive}
\end{align} 
the variation of $S_0$ with respect to $\Psi$ would have given
\begin{subequations}
\begin{align}
\delta S_0&=\omega_M(\delta\Psi, \hat{Q}_\mathrm{B}\Psi)-\frac12\omega(\delta\Psi, \hat{B}\Psi)\\
&=\int_M d^d x \,\omega'\big(\delta\Psi(x),Q_\mathrm{B}\Psi(x)\big)-\frac{1}{2}\int_{\partial M} d^{d-1}y \,\sqrt{\gamma}\,n_\mu(y) \, \omega'\big(\Psi(y), \alpha' c \partial^\mu\delta\Psi(y)\big)+\nonumber\\
&\hspace{3.8cm}+\frac{1}{2}\int_{\partial M} d^{d-1}y \,\sqrt{\gamma}\,n_\mu(y)\,  \omega'\big(\delta\Psi(y), ( \alpha' c \partial^\mu+\Omega^\mu)\Psi(y)\big)\,,
\end{align}
\end{subequations}
where we have used the shortcut $\Psi(y^a)\equiv \Psi(x^\mu(y^a))$.
The second term in the variation can be made vanishing by imposing Dirichlet boundary condition
\begin{align}
\delta\Psi(x)\big|_{\partial M}=0\,,
\end{align}
or the Neumann boundary condition
\begin{align}
n_\mu(\alpha' c\partial^\mu+\Omega^\mu)\Psi(x)\big|_{\partial M}=0\,.\label{B-neum}
\end{align}
However, forcing the first term to be zero (after having fixed Neumann or Dirichlet conditions) would generally over-constrain the system. The variational principle for the field theory on $M$ given by the action \eqref{eq:actionNaive} is therefore ill-defined. In addition to this deficiency, specializing to the gauge variation $\delta_\Lambda\Psi = \hat{Q}_\mathrm{B}\Lambda$, we find that 
\begin{align}
    \delta_\Lambda S_0 &=-\frac{1}{2}\omega( \Lambda,\hat{Q}_\mathrm{B}\hat{B}\Psi)\,.\label{eq:gaugeS0}
\end{align}
Since the product $\hat{Q}_\mathrm{B}\hat{B}$ is generally non-vanishing, we conclude that for general gauge parameter $\Lambda$, the r.h.s.\ of \eqref{eq:gaugeS0} will be non-zero.
This leaves us with the conclusion that the action $S_0$ can, at the very best, be invariant under a restricted class of bulk gauge transformations which are given by such $\Lambda$ for which the r.h.s.\ of \eqref{eq:gaugeS0} vanishes for all $\Psi$.

It will be the subject of the remainder of this paper to attempt to fix both the well-definedness of the variational principle given by $S_0$ as well as its gauge-invariance by adding suitable boundary terms. That is, we will seek a total action
\begin{align}
    S= S_0 + S_{\p M}\,,
\end{align} 
where the terms $S_{\p M}$ localize on $\p M$, such that the action $S$
\begin{enumerate}
\itemsep0em 
    \item gives a well-defined variational principle,
    \item is invariant w.r.t.\ the gauge transformation $\delta_\Lambda \Psi = \hat{Q}_\mathrm{B}\Psi$ of the target fields $\Psi_s(x)$,
    \item reduces to $S_0$ upon assuming suitable fall-off conditions on $\Psi_s(x)$ towards $\p M$.
\end{enumerate}

\subsubsection{Restoring cyclicity of the kinetic operator}
\label{subsec:restoring_cyc}

Let us first attempt to construct a boundary term which would make the variational principle well-defined.
To this end it is convenient to generally write the boundary operator $B$ in the form
\begin{align}
    B = \delta(x)\,\Gamma^\ast + \Gamma\, \delta(x)\,, \label{eq:Bdecomp}
\end{align}
where 
\begin{align}
    \Gamma^\ast = -\alpha' c\, n^\mu(y)\,\p_\mu -\gamma^\ast\label{eq:GammaDecomp}
\end{align}
and $\gamma^*$ is any operator which is free of target derivatives and which satisfies
\begin{align}
    \gamma^* +  \gamma= \Omega^\mu n_\mu(y)\,.\label{eq:gammaDecomp}
\end{align}
Notice that this fixes $\gamma^*$ only up to possible BPZ-odd corrections. For example, an immediate simple choice (which effectively reproduces the results of  \cite{georg} and \cite{atakan}) would have been
\begin{align}
\gamma^*_\text{simple}=\frac{1}{2}\Omega^\mu n_\mu(y)\,.\label{G-trivial}
\end{align}
Given the above definitions, let us consider adding the boundary term
\begin{align}
    S_{\p M,\text{cyc}}(\Psi) = -\frac{1}{2}\omega(\Psi,\delta(\hat{x})\hat{\Gamma}^*\Psi)\label{eq:ScycBndy}
\end{align}
to the bulk action $S_0$, thus obtaining the total bulk plus boundary action of the form
\begin{align}
S_\mathrm{cyc}(\Psi)=S_0(\Psi)+S_{\p M,\text{cyc}}(\Psi)=\frac12\omega\big(\Psi,(\theta(\hat{x}) \hat{Q}_\mathrm{B}-\delta(\hat{x})\hat{\Gamma}^*)\Psi\big)\,.\label{Stot}
\end{align}
Notice that the total kinetic operator $\theta(\hat{x}) \hat{Q}_\mathrm{B}-\delta(\hat{x})\hat{\Gamma}^*$ of $S_\mathrm{cyc}$ is now BPZ odd. Indeed, using the commutation relation \eqref{eq:Bdef} and the decomposition \eqref{eq:Bdecomp} of $\hat{B}$, we can write
\begin{subequations}
\label{eq:bpz_proof}
\begin{align}
\big(\theta(\hat{x}) \hat{Q}_\mathrm{B}-\delta(\hat{x})\hat{\Gamma}^*\big)^*&=-\hat{Q}_\mathrm{B}\theta(\hat{x})-\hat{\Gamma}\delta(\hat{x})\\
&=-\theta(\hat{x}) \hat{Q}_\mathrm{B}+(\delta(\hat{x})\hat{\Gamma}^\ast + \hat{\Gamma} \delta(\hat{x}))-\hat{\Gamma}\delta(\hat{x})\\
&=-\big(\theta(\hat{x}) \hat{Q}_\mathrm{B}-\delta(\hat{x})\hat{\Gamma}^\ast\big)\,.
\end{align}
\end{subequations}
This enables us to write the variation of the total action $S_\mathrm{cyc}$ as
\begin{subequations}
    \begin{align}
        \delta S_\mathrm{cyc}&=\omega(\delta\Psi,(\theta(\hat{x}) \hat{Q}_\mathrm{B}-\delta(\hat{x})\hat{\Gamma}^*)\Psi)\,,\\[1mm]
        &= \int_{M}d^d x\, \omega'\big(\delta\Psi(x),Q_\mathrm{B}\Psi(x)\big)+\nonumber\\
        &\hspace{2cm}+\int_{\partial M}d^{d-1} y\,\sqrt{\gamma}\, \omega'\big(\delta\Psi(y),(\alpha' c\, n^\mu(y)\,\p_\mu +\gamma^\ast)\Psi(y)\big)\,,\label{eq:varCyc}
    \end{align}
\end{subequations}
where the entire boundary contribution can now be set to zero by imposing either the Dirichlet boundary condition
\begin{align}
\delta\Psi(x)\big|_{\partial M}=0\,,
\end{align}
or the Neumann boundary condition
\begin{align}
\Gamma^\ast\Psi(x)\big|_{\partial M}=-(\alpha' cn^\mu \partial_\mu+\gamma^\ast)\Psi(x)\big|_{\partial M}=0\,.
\end{align}
This confirms that the action $S_\mathrm{cyc}$ given by \eqref{Stot} gives a well-defined variational problem. Another way of appreciating this fact is to rewrite the combined action \eqref{Stot} in terms of an explicit target integral using the derivative decompositions \eqref{eq:Qdecomp} of $Q_\mathrm{B}$ and \eqref{eq:GammaDecomp} of $\Gamma^\ast$. Integrating by parts the bulk term containing second derivative with respect to target coordinates and using the divergence theorem, we obtain
\begin{align}
    S_\mathrm{cyc}(\Psi)&=\frac{1}{2}\int_{M}d^d x\, \omega'\big(\p_\mu\Psi(x),\alpha' c\p^\mu\Psi(x)\big)+\frac{1}{2} \int_{M}d^d x\, \omega'\big(\Psi(x),(Q'-\Omega^\mu\p_\mu)\Psi(x)\big)+\nonumber\\
    &\hspace{6.6cm}+\frac{1}{2}\int_{\p M}d^{d-1} y\, \sqrt{\gamma}\,\omega'\big(\Psi(y),\gamma^\ast \Psi(y)\big)\,.\label{eq:explicitS}
\end{align}
This is an action which 1.\ contains no second derivatives in the bulk terms, 2.\ is free of normal derivatives in the boundary terms. As such, it is guaranteed to give a well-defined variational principle. It is not hard to check that the variation of the explicit form \eqref{eq:explicitS} of the action agrees with the result \eqref{eq:varCyc}.

Second, specializing to the gauge variation and using the nilpotency of $\hat{Q}_\mathrm{B}$, we end up with a pure-boundary term 
\begin{align}
    \delta_\Lambda S_\mathrm{cyc} = -\int_{\p M}d^{d-1} y\, \sqrt{\gamma}\,\omega'\big(\Psi(y),\Gamma^\ast {Q}_\mathrm{B}\Lambda(y)\big)\,.\label{eq:gt_cyc}
\end{align}
Hence, if the action $S_\mathrm{cyc}$ is to be fully gauge invariant, this should be vanishing for a generic ghost-number one string field $\Psi$ and generic ghost-number zero gauge parameter $\Lambda$. However, even after taking into account the aforementioned freedom in the definition of $\gamma^\ast$, it seems generally impossible to ensure that $\Gamma^\ast {Q}_\mathrm{B}=0$ at the operatorial level. Hence, we are left with the conclusion that, although the class of boundary terms $S_\mathrm{\p M,\mathrm{cyc}}$ described by \eqref{eq:ScycBndy} rectifies the problem with well-definedness of the variational principle, it does not seem to be rich enough to make the string field theory action defined on $M$ invariant under gauge transformations with respect to an unrestricted gauge parameter $\Lambda$.

On the positive side, one can note that the action $S_\mathrm{cyc}$ provides a good action on $M$ for both the open-, as well as for the closed-string tachyon. These appear at level $-1$, where there is no non-trivial gauge parameter.



\subsubsection{Preliminary comments on restoring gauge invariance}
\label{subsec:comments}

In order to get some ideas of what steps need to be taken in order to make the string field theory action on a target $M\subset \mathbb{R}^d$ fully gauge-invariant, let us try to learn some lessons by considering examples of simpler field theories.

Let us start by examining general relativity on a manifold $M$ with a boundary $\p M$. Its variational principle is given by the action
\begin{align}
    S_\mathrm{GR}= S_\mathrm{EH}+S_\mathrm{GHY}= \int_M d^d x\,\sqrt{g}\, R +\int_{\p M} d^{d-1} y\,\sqrt{\gamma}\, 2K\,,\label{eq:actionGR}
\end{align}
where $R$ is the Ricci scalar and $K$ is the trace of the extrinsic curvature tensor.
Here the usual Einstein-Hilbert term is supplemented by the Gibbons-Hawking-York boundary term which, upon varying, cancels other boundary terms containing normal derivatives of the metric variation produced by the Einstein-Hilbert term. Expanding this action in small fluctuations $h_{\mu\nu}$ of the metric around a background configuration $\overline{g}_{\mu\nu}$, one should certainly hope to obtain a consistent action for massless spin-2 fluctuations on $M$, which is invariant under the gauge transformation
\begin{align}
    \delta_\xi h_{\mu\nu} = \overline{\nabla}_\mu\xi_\nu +\overline{\nabla}_\nu\xi_\mu\,.\label{eq:gaugeGrav}
\end{align}
Here $\overline{\nabla}_\mu$ denotes the covariant derivative defined using the Levi-Civita connection associated with the background metric $\overline{g}_{\mu\nu}$ around which we are expanding. Checking the transformation of such a linearized action under \eqref{eq:gaugeGrav}, one quickly realizes that it is invariant only after restricting the gauge parameter $\xi_\mu$ to satisfy $n^\mu\xi_\mu=0$ at $\p M$. Of course, this was to be expected, since the gauge transformation \eqref{eq:gaugeGrav} is inherited from the diffeomorphism invariance of the action \eqref{eq:actionGR} and, assuming that $\p M$ is fixed, the action \eqref{eq:actionGR} is only invariant under diffeomorphisms which preserve $\p M$. At the same time, this tells us that if we wish to obtain an action for the massless spin-2 fluctuation $h_{\mu\nu}$ on $M$ which would be invariant under \eqref{eq:gaugeGrav} with an \emph{unconstrained} gauge parameter $\xi_\mu$, we have to allow for the boundary $\p M$ to fluctuate. This has the effect of introducing an {additional degree of freedom} into the linearized action, which is associated with the transverse displacement of the boundary and which therefore only contributes to the action through boundary terms. When this degree of freedom is endowed with an appropriate gauge-transformation law (proportional to the component of the infinitesimal diffeomorphism $\xi_\mu$ normal to the boundary), invariance of the linearized action under the full gauge group of massless spin-2 is restored. Thus, realizing that string field theory on a flat non-compact background generally contains a massless spin-2 sector (level 0 of the closed string and level 1 of the open string in the tensionless limit), we can make the following expectation: 

\begin{center}
    \emph{A fully gauge-invariant action for string field theory on $M\subset \mathbb{R}^d$ may feature additional degrees of freedom which only enter the action through the boundary terms $S_{\p}$.}
\end{center}

Second, expanding the action \eqref{eq:actionGR} in small fluctuations $h_{\mu\nu}$ of the bulk metric, one generally first encounters both bulk and boundary terms which are linear in $h_{\mu\nu}$: the tadpoles. Since the bulk tadpoles 
\begin{align}
    \int_M d^d x \,\sqrt{\overline{g}}\,\big(\tfrac{1}{2}\overline{R}\overline{g}^{\mu\nu}-\overline{R}^{\mu\nu}\big)\,h_{\mu\nu}
\end{align}
are proportional to the Einstein field equations for the background metric, they can be set to zero by assuming that the bulk, around which we are perturbing, is critical. For instance, this is certainly true when $M\subset \mathbb{R}^d$. The vanishing of the boundary tadpoles\footnote{Here we define $\gamma^{\mu\nu} = g^{\mu\nu}-n^\mu n^\nu = \tensor{e}{^\mu_a}\tensor{e}{^\nu_b}\gamma^{ab}$ where $\gamma^{ab}$ is the inverse of the induced metric $\gamma_{ab} = \tensor{e}{^\mu_a}\tensor{e}{^\nu_b} g_{\mu\nu}$, on the boundary. We also define $\tensor{e}{^\mu_a}=\p x^\mu(y)/\p y^a$ where $x^\mu(y)$ is the embedding map defining the hypersurface $\p M$, and the extrinsic curvature $K_{\mu\nu} = \nabla_\mu n_\nu - n_\mu n^\lambda\nabla_\lambda n_\nu$. Taking $\p M$ to be rigid (fixing the embedding maps $x^\mu(y)$ under variations), the boundary tadpole \eqref{eq:bndy_tadpole} can be alternatively recast as
\begin{align}
    \int_{\p M} d^{d-1} y \,\sqrt{\overline{\gamma}}\,\big(\overline{K}\overline{\gamma}^{ab}-\overline{K}^{ab}\big)\,\delta\gamma_{ab}= -\int_{\p M} d^{d-1} y \,\sqrt{\overline{\gamma}}\,\big(\overline{K}\overline{\gamma}_{ab}-\overline{K}_{ab}\big)\,\delta\gamma^{ab}\,,
\end{align}
where ${K}_{ab} = \tensor{e}{^\mu_a}\tensor{e}{^\nu_b} K_{\mu\nu}$ (or, equivalently, ${K}^{\mu\nu}=\tensor{e}{^\mu_a}\tensor{e}{^\nu_b}{K}^{ab}$) and $\delta\gamma_{ab} = \tensor{e}{^\mu_a}\tensor{e}{^\nu_b} h_{\mu\nu}$.
See Appendix \ref{app:linear} for more details on the notation.
} 
\begin{align}
    \int_{\p M} d^{d-1} y \,\sqrt{\overline{\gamma}}\,\big(\overline{K}\overline{\gamma}^{\mu\nu}-\overline{K}^{\mu\nu}\big)\,h_{\mu\nu}\label{eq:bndy_tadpole}
\end{align}
can in turn be ensured either 1.\ by requiring that the fluctuations $h_{\mu\nu}$ respect suitable fall-off conditions as we approach $\p M$ (as would be appropriate if the classical background, around which we are expanding, was endowed with Dirichlet boundary conditions at $\p M$), or, 2.\ by requiring that the background $(M,\overline{g}_{\mu\nu})$ is chosen so that the extrinsic curvature $\overline{K}^{\mu\nu}$ of $\p M$ satisfies the (Neumann) condition
\begin{align}
    \overline{\gamma}^{\mu\nu} \overline{K} -\overline{K}^{\mu\nu}  =0\,.\label{eq:bndyEOMGR}
\end{align}
This will not hold for a generic $\p M$, even when $M\subset \mathbb{R}^d$ is flat. As a consequence, one should expect that criticality of an action on $M$ for massless spin-2 fluctuations $h_{\mu\nu}$ for which we would \emph{not} insist on any fall-off conditions on $h_{\mu\nu}$ towards $\p M$, would generally be spoiled by the presence of boundary tadpoles, even if the bulk is critical. It is only for the special cases of backgrounds which obey \eqref{eq:bndyEOMGR} (such as a flat boundary in flat $\mathbb{R}^d$) that one can achieve full criticality both in the bulk and on the boundary without imposing any fall-off on $h_{\mu\nu}$ towards $\p M$. This leads us to our second expectation:
\begin{center}
    \emph{
    A generic choice of $M \subset \mathbb{R}^d$ bounded by a generic curved boundary $\p M$ may result in the presence of boundary tadpoles in the string field theory action.}
\end{center}
Unfortunately, unlike for pure gravity, for string field theory we do not seem to be lucky enough to have a simple ``background-independent'' formulation at our disposal. Correspondingly, there does not seem to be a simple way of telling what fall-off conditions should be assumed for a particular boundary $\p M$ of $M \subset \mathbb{R}^d$, or, what shapes (if any) of $\p M$ stand a chance of furnishing a background for string field theory which would be free of boundary tadpoles with no a priori need for any fall-off conditions. At the same time, attempting to formulate a consistent action for an off-shell string field theory with boundary tadpoles appears to be a formidable task: in order to meaningfully expand an action around a point which is not a true classical vacuum of the theory, one would have to take into account all interactions. 

With all this in mind, we will spend the rest of this paper by trying validate at least one example of a background with a boundary, where one can construct a tadpole-free gauge-invariant action for string fields in the bulk of $M$ \emph{without} assuming any fall-off conditions towards $\p M$. 
In particular, motivated by the discussion in this section, we will be considering a \emph{flat} boundary $\p M$ bounding a subset $M$ of a flat euclidean space $\mathbb{R}^d$. That is, the case when $M$ is a half-space and $\p M$ the $(d-1)$-dimensional hyperplane which is bounding it. 

\section{Massless level with a flat boundary}\label{sec:massless}

Since the tachyon action, which generally arises at level $-1$ of string field theory, is \emph{not} constrained by gauge invariance, it is the massless fields arising at level 0 which provide the simplest nontrivial proving ground for any construction of a would-be gauge-invariant action on $M\subset \mathbb{R}^d$ with a boundary $\p M$. Taking motivation from our general discussion in Section \ref{subsec:comments}, we will focus on the simple case where $M$ is taken to be a half-space bounded by a codimension-one hyperplane $\p M$.\footnote{See \cite{Ahmadain:2024uyo} for recent results on string theory on the same half-space in relation to the GHY boundary term, using the world-sheet sigma-model approach.} After outlining in Sections \ref{subsec:init} and \ref{subsec:general} a general strategy for constructing a gauge-invariant action for free massless string fields on such a background, we will obtain in Sections \ref{subsec:open} and \ref{subsec:closed} explicit forms of the corresponding actions for the target fluctuations which arise at level 0 of both the open-string and the closed-string spectrum.

\subsection{Initial setup}
\label{subsec:init}

For the sake of streamlining the discussion, let us from now on assume that there are no compact internal sectors and that the ghost central charge is fully balanced by the $d=26$ non-compact free bosons. Our results can then be extended to more complicated setups more-or-less straightforwardly. To be concrete, pick a coordinate system 
$$
x^{\mu}=(y^a,z)\,,
$$ 
where $y^a$ are Cartesian coordinates along $\p M$ and the $z$-axis is normal to $\p M$. The Greek indices $\mu,\nu,\ldots$ will run over the $d$ directions of $M$ while the Latin indices $a,b,\ldots$ will be confined to the $(d-1)$-dimensional hyperplane $\p M$. The boundary $\p M$ will be chosen to sit at $z=0$ while the negative $z$-axis $z<0$ extends into the bulk of the half-space $M$. Moreover, since the boundary $\p M$ is assumed to be flat, its unit normal is given by a constant vector field $n^\mu (y)=n^\mu$. As such, it commutes with all target derivatives -- a property which will be instrumental in our construction below. In the above-described coordinates, $n^\mu$ reads simply $(0,\ldots,0,1)=\delta^\mu_z$.

\subsubsection{Field content}

Generic massless components of the full string field $\Psi(\hat{x})|0\rangle$ for the open and closed string, respectively, propagating in $d=26$ non-compact dimensions, are given by 
\begin{align}\label{eq:masslessfields}
\begin{array}{lr}
     \Big[A_\mu(\hat{x}) \,c_1 \alpha_{-1}^\mu - i\sqrt{\frac{\alpha'}{2}} B(\hat{x}) \, c_0 \Big] |0\rangle\,, &  \hspace{4mm}\text{(open string)}\\[7mm]
      \Big[-\frac{1}{2}e_{\mu\nu}(\hat{x}) \alpha_{-1}^\mu \bar{\alpha}_{-1}^\nu c_1\bar{c}_1+e(\hat{x})c_1c_{-1}+\bar{e}(\hat{x})\bar{c}_1\bar{c}_{-1}+ & \\[1mm]
    \hspace{3cm}+i\sqrt{\frac{\alpha'}{2}} c_0^+ \big( e_\mu(\hat{x}) c_1\alpha_{-1}^\mu+\bar{e}_\mu(\hat{x}) \bar{c}_1\bar{\alpha}_{-1}^\mu\big)\Big]|0\rangle\,. &  \hspace{4mm}\text{(closed string)}\\
\end{array}
\end{align}
These give rise to the target fields 
\begin{align}
    A_\mu(x)\,,\quad B(x)
\end{align}
of the open string theory (the gauge-field and the auxiliary Nakanishi-Lautrup field) and the target fields
\begin{align}
    e_{\mu\nu}(x)\,,\quad e_\mu(x)\,,\quad \bar{e}_\mu(x)\,,\quad e(x)\,,\quad \bar{e}(x)
\end{align}
of the closed string theory (these will be later seen to reorganize themselves into the graviton, the Kalb-Ramond field, dilaton and two auxilliary vector fields).

\subsubsection{Truncated BRST operator}

Acting on such states, the full BRST charge can be truncated down to terms containing only the oscillators $c_0$, $b_0$, $\alpha_{1}^\mu$, $\alpha_{- 1}^\mu$, $c_{1}$, $c_{-1}$ for the open string and $c_0^+$, $b_0^+$, $\alpha_{1}^\mu$, $\alpha_{-1}^\mu$, $c_{1}$, $c_{-1}$, $\bar{\alpha}_{1}^\mu$, $\bar{\alpha}_{- 1}^\mu$, $\bar{c}_{1}$, $\bar{c}_{-1}$ for the closed string. In particular, following the structure of the target-derivative decomposition \eqref{eq:Qdecomp} of $Q_\mathrm{B}$, one can explicitly identify
\begin{align}
\label{eq:openQ}
    c &= c_0\,,\nonumber\\
    \Omega^\mu &= i\sqrt{2\alpha'}\big(c_1 \alpha_{-1}^\mu+c_{-1}\alpha_1^\mu\big) \,,\hspace{4.8cm} \text{(open string)}\\
    Q'&=-2b_0 c_{-1}c_{1}\,,\nonumber
\end{align}
for the open string, as well as
\begin{align}
\label{eq:closedQ}
    c &= \frac{1}{2}c_0^+\,,\nonumber\\
    \Omega^\mu &= i\sqrt{\frac{\alpha'}{2}}\big(c_1 \alpha_{-1}^\mu+c_{-1}\alpha_1^\mu+\bar{c}_1 \bar{\alpha}_{-1}^\mu+\bar{c}_{-1}\bar{\alpha}_1^\mu\big) \,,\hspace{1.8cm} \text{(closed string)}\\[2mm]
    Q'&=-b_0^+ \big(c_{-1}c_{1}+\bar{c}_{-1}\bar{c}_1\big)\,.\nonumber
\end{align}
for the closed string. Given these definitions and the standard commutation relations for the matter and ghost oscillators, it is straightforward to check that the operators $c$, $\Omega^\mu$ and $Q'$ defined by \eqref{eq:openQ} and \eqref{eq:closedQ} indeed satisfy the algebra \eqref{eq:Qalgebra}. This, in turn, guarantees the nilpotency of $Q_\mathrm{B}$.

\subsection{Algebraic framework for $S_{\p M}$}
\label{subsec:general}

Let us now lay out a construction of a boundary term $S_{\p M}$ for the massless part of the SFT action on a subset $M$ of $\mathbb{R}^d$ with a flat boundary, which \emph{both} restores the well-definedness of the variational principle, as well as makes the action gauge-invariant. We will keep the discussion as general as possible in order to accommodate both the open-string, as well as the closed-string theory at once. Explicit expressions for the target actions of these two theories at level 0 will then be worked out below in Sections \ref{subsec:open} and \ref{subsec:closed}, respectively.

\subsubsection{Mode decomposition of $\Omega^\mu$}

One can first notice that the operator $\Omega^\mu$ can be naturally decomposed as
\begin{align}
    \Omega^\mu = \Omega^\mu_++\Omega^\mu_-\,,\label{eq:Omegasplit}
\end{align}
where $\Omega_+^\mu$ is understood to contain only the positively-moded matter oscillators while $\Omega_-^\mu$ only the negatively-moded ones. Looking at the concrete realizations \eqref{eq:openQ} and \eqref{eq:closedQ} of $\Omega^\mu$, it follows that
\begin{align}
    (\Omega_{\pm}^\mu)^\ast = \Omega_{\mp}^\mu\,,\label{eq:BPZOm}
\end{align}
as well as that
\begin{subequations}
\begin{align}
    [c,\Omega_{\pm}^\mu]&=0\,,\label{eq:Omc}\\
    [\Omega_{\pm}^\mu,\Omega_{\pm}^\nu]&=0\,,\label{eq:Omnilp}\\
    -\alpha'\eta^{\mu\nu}[c,Q']+[\Omega_{+}^\mu,\Omega_{-}^\nu]&=0\,.
\end{align}
\end{subequations}
In particular, the BPZ property \eqref{eq:BPZOm} implies that the choice $\gamma^\ast= n_\mu\Omega_+^\mu\equiv \Omega_+^z$ for the operator $\gamma^\ast$ introduced in Section \ref{subsec:restoring_cyc} ensures validity of the requirement \eqref{eq:gammaDecomp}. Hence, substituting into the general formula \eqref{Stot} with the corresponding choice 
\begin{align}
\Gamma^* = -\alpha' c\, \p_z -\Omega_+^z\label{eq:Gamma_star_plus}
\end{align}
of the operator $\Gamma^*$, one arrives at an action
\begin{align}
    S_{\mathrm{cyc},+}(\Psi)
    &=\frac{1}{2}\int_{M}d^d x\, \omega'\big(\p_\mu\Psi(x),\alpha' c\p^\mu\Psi(x)\big)+\nonumber\\
    &\hspace{4.8cm}+\frac{1}{2} \int_{M}d^d x\, \omega'\big(\Psi(x),(Q'-2\Omega_-^\mu\p_\mu)\Psi(x)\big)\,.\label{eq:Scyc+}
\end{align}
Although this represents a well-defined variational problem, as discussed above, it generally fails to be gauge-invariant.

\subsubsection{Boundary string fields}
\label{subsec:boundary_sf}

At this point, in agreement with our expectations which we have outlined in Section \ref{subsec:comments}, let us consider introducing two additional level-0 string fields
\begin{align}
    |\Sigma_0\rangle \equiv \Sigma_0(\hat{y})|0\rangle\,,\qquad |\Sigma_1\rangle \equiv \Sigma_1(\hat{y})|0\rangle\,,
\end{align}
which, by definition, are localized on $\p M$ and which are otherwise taken to be linear combinations of the same basis states as the bulk string field $\Psi(x)$. We will choose to endow $\Sigma_0$ and $\Sigma_1$ with the gauge-transformation laws
\begin{subequations}
\label{eq:bndy_gauge_tr}
    \begin{align}
        \delta_\Lambda \Sigma_0(y) &= \alpha' c \Lambda^{(1)}(y)+\Omega_+^z\Lambda^{(0)}(y)\,,\\[1.0mm]
        \delta_\Lambda \Sigma_1(y) &= \alpha' c \Lambda^{(2)}(y)+\Omega_+^z\Lambda^{(1)}(y)\,.
    \end{align}
\end{subequations}
Here we have denoted
\begin{align}
    \Lambda^{(k)}(y) \equiv \p_z^k \Lambda(x)|_{z=0}
\end{align}
where $\p_z$ is the derivative normal to $\p M$ and $\Lambda(x)$ is the \emph{same} gauge parameter as the one which is used to transform the bulk string field $\Psi(x)$. For the moment, we can take \eqref{eq:bndy_gauge_tr} to be a judicious choice, which, in the end, will lead to a gauge-invariant action. 

Generally speaking, we will be aiming to extend the bulk action and the spectrum of bulk degrees of freedom in a minimal way which restores gauge invariance of the theory. To this end, we will always have in mind making as economic a choice of the states entering $\Sigma_0$ and $\Sigma_1$ as possible, which, nevertheless, is still consistent with the gauge transformation \eqref{eq:bndy_gauge_tr}. In particular, we note that one can consistently truncate the boundary string fields $\Sigma_0(\hat{y})$ and $\Sigma_1(\hat{y})$ by requiring that they satisfy the linear relation
\begin{align}
    \alpha' c \,\Sigma_1 +\Omega_+^z \Sigma_0=0\,.\label{eq:lin_rel}
\end{align}
Indeed, making use of the properties \eqref{eq:Omc} and \eqref{eq:Omnilp}, one can quickly verify that the relation \eqref{eq:lin_rel} is invariant under the transformation of $\Sigma_0$ and $\Sigma_1$ dictated by \eqref{eq:bndy_gauge_tr}. Moreover, since the l.h.s.\ of \eqref{eq:lin_rel} is purely algebraic (meaning that it does not contain any target derivatives), it can be solved rather easily, as soon as we substitute for explicit expansions of $\Sigma_0$ and $\Sigma_1$ in terms of linear combinations of the appropriate level-0 states. We will see this procedure at work later in Sections \ref{subsec:open} and \ref{subsec:closed}.

\subsubsection{Flat-boundary decomposition of ${Q}_\mathrm{B}$}

Recall that when introducing the decomposition \eqref{eq:Qdecomp}, we chose to make explicit the dependence of ${Q}_\mathrm{B}$ on target derivatives in order to demonstrate that the naive kinetic term \eqref{eq:actionNaive} does not represent a good action on a domain with a boundary.
In the case of a flat boundary $\p M$, we will find it beneficial to pull out only the ``dangerous'' derivatives $\p_z$ with respect to the coordinate normal to $\p M$, which cannot be integrated by parts without producing boundary terms. This leads us to the decomposition
\begin{align}
    Q_\mathrm{B} =-\alpha' c \p_z^2-\Omega^z \p_z +\tilde{Q}\,,
\end{align}
where $\tilde{Q}$ now contains first and second derivatives with respect to the coordinates $y^a$ along the boundary. These, however, \emph{can} be integrated by parts, by assumption. The operator $\tilde{Q}$ is therefore cyclic with respect to $\omega_M$. More explicitly, one can write
\begin{align}
    \tilde{Q} = -\alpha' c \p_a^2-\Omega^a \p_a +Q'\,.
\end{align}
In addition, we can note that the nilpotency of $Q_\mathrm{B}$ is equivalent to the algebra
\begin{subequations}
   \begin{align}
    c^2&=0\,,\\
    [c,\Omega^z]&=0\,,\label{eq:cOmz}\\
    -\alpha'[c,\tilde{Q}]+\tfrac{1}{2}[\Omega^z,\Omega^z]&=0\,,\\
    [\tilde{Q},\Omega^z]&=0\,,\label{eq:tQz}\\
    \tilde{Q}^2&=0\,.
\end{align}
\end{subequations}
This, in particular, implies that $\tilde{Q}$ is also nilpotent. It therefore seems natural to think of $\tilde{Q}$ as the reduction of $Q_\mathrm{B}$ to $\p M$. In addition, since $\Omega^z$ is assumed to be free of zero modes, note that the relation \eqref{eq:cOmz} generally holds separately for $\Omega_+^z$ and $\Omega_-^z$.

We will soon see that for writing down a gauge-invariant action, it is key to realize that, at the massless level, the operators $\Omega_+^z$ and $\Omega_-^z$ both separately anti-commute with $\tilde{Q}$, namely
\begin{align}
    [\tilde{Q},\Omega_{\pm}^z]=0\,,\label{eq:SepComm}
\end{align}
whereas the nilpotency of $Q_\mathrm{B}$ only guarantees a somewhat weaker relation \eqref{eq:tQz}. 

\subsubsection{Gauge-invariant action}

We are now ready to write down an action for $\Psi(x)$, $\Sigma_0(y)$, $\Sigma_1(y)$ which is invariant under the gauge transformations \eqref{eq:bndy_gauge_tr} of the boundary string fields and, crucially, under the gauge transformation $\delta_\Lambda \Psi = \hat{Q}_\mathrm{B}\Lambda$ of the bulk string field. It reads
\begin{subequations}
\label{eq:TheAction}
\begin{align}
    &S(\Psi;\Sigma_0,\Sigma_1) =\nonumber\\ 
     &\hspace{0.8cm}=S_{\mathrm{cyc},+}(\Psi)+\int_{\p M}d^{d-1} y \,\omega'\big(\Psi+\Sigma_1, \tilde{Q}\Sigma_0-\Omega_-^z \Sigma_1\big)\label{eq:TheAction1}\\
     &\hspace{0.8cm}=\frac{1}{2}\int_{M}  d^d x \,\omega'\big(\p_\mu\Psi(x), c\p^\mu\Psi(x)\big)+\frac{1}{2} \int_{M}  d^d x\, \omega'\big(\Psi(x),(Q'- 2\Omega_-^\mu\p_\mu)\Psi(x)\big)+\nonumber\\
    &\hspace{5.0cm}+\int_{\p M}  d^{d-1} y \,\omega'\big(\Psi(y)+\Sigma_1(y), \tilde{Q}\Sigma_0(y)-\Omega_-^z \Sigma_1(y)\big)\,,
\end{align}
\end{subequations}
where the boundary string fields $\Sigma_0$ and $\Sigma_1$ are assumed to safisfy the linear relation \eqref{eq:lin_rel} and where $S_{\mathrm{cyc},+}(\Psi)$ was already constructed in \eqref{eq:Scyc+} so as to give a well-defined variational principle for $\Psi$. 

Since the second term on the r.h.s.\ of \eqref{eq:TheAction1} does not contain any derivatives normal to $\p M$, and, since the tangential derivatives $\p_a$ contained in $\tilde{Q}$ can be integrated by parts without producing corner terms, it follows that the action $S(\Psi;\Sigma_0,\Sigma_1)$ given by \eqref{eq:TheAction} represents a well-defined variational principle for $\Psi$, $\Sigma_0$ and $\Sigma_1$. 

It remains to check the properties of \eqref{eq:TheAction} under the gauge transformation of $\Psi$, $\Sigma_0$ and $\Sigma_1$. Performing first the gauge variation of $S_{\mathrm{cyc},+}(\Psi)$, we can establish that 
    \begin{align}
        \delta_\Lambda S_{\mathrm{cyc},+}(\Psi) &=-\int_{\p M} d^{d-1} y \,\omega'\big(\Psi,\tilde{Q}\delta_\Lambda\Sigma_0-\Omega_-^z \delta_\Lambda\Sigma_1\big) \,.\label{eq:dScyc}
    \end{align}
To arrive at this result, we have recalled the calculation \eqref{eq:gt_cyc} and noted that using the various algebraic properties outlined above (including \eqref{eq:SepComm}), we can rewrite
\begin{align}
    \Gamma^\ast Q_\mathrm{B} = -\Omega_-^z\big(\alpha' c\p_z^2 +\Omega_+^z \p_z\big)+\tilde{Q}\big(\alpha' c\p_z +\Omega_+^z \big)\,,
\end{align}
where on the right-hand side, we recognize the differential operators defining the gauge transformation \eqref{eq:gt_cyc} of $\Sigma_1$ and $\Sigma_0$.
On the other hand, applying the gauge transformation to the second term on the r.h.s.\ of \eqref{eq:TheAction1} in turn gives
\begin{align}
   \hspace{-1mm} \int_{\p M}d^{d-1} y \,\omega'\big(\Psi, \tilde{Q}\delta_\Lambda\Sigma_0-\Omega_-^z \delta_\Lambda\Sigma_1\big)+\int_{\p M}d^{d-1} y \,\omega'\big(\Omega_-^z\Lambda^{(2)}\!-\!\tilde{Q}\Lambda^{(1)}, \alpha' c \,\Sigma_1 +\Omega_+^z \Sigma_0\big)\,,\label{eq:var2}
\end{align}
where the first term exactly cancels with the r.h.s.\ of \eqref{eq:dScyc}, while the second term vanishes as a consequence of the assumed linear relation \eqref{eq:lin_rel} between $\Sigma_0$ and $\Sigma_1$. Thus, in total, we conclude that
\begin{align}
    \delta_\Lambda S(\Psi;\Sigma_0,\Sigma_1) =0\,,
\end{align}
meaning that \eqref{eq:TheAction} defines a gauge-invariant action. As we discuss in detail in Appendix \ref{app:nilpotent}, this gauge-invariance arises as a consequence of an underlying cyclic nilpotent structure.

Finally, notice that the structure of the variation \eqref{eq:var2} suggests that,
by integrating in one more boundary string field $\Xi(y)$, which transforms under the gauge transformation as
\begin{align}
    \delta_\Lambda \Xi(y) = -\Omega_-^z\Lambda^{(2)}+\tilde{Q}\Lambda^{(1)}\,,
\end{align}
one can write down an action 
\begin{align}
    \mathcal{S}(\Psi;\Sigma_0,\Sigma_1,\Xi) = S(\Psi;\Sigma_0,\Sigma_1)+\int_{\p M}d^{d-1} y \,\omega'\big(\Xi, \alpha' c \,\Sigma_1 +\Omega_+^z \Sigma_0\big)\,,\label{eq:curlyS}
\end{align}
which is gauge-invariant off-shell \emph{without} assuming any relation between the boundary string fields $\Sigma_0$, $\Sigma_1$, $\Xi$. Indeed, when one checks the gauge transformation of the new term in \eqref{eq:curlyS}, substituting the gauge variation of $\Xi$ has the effect of canceling the second term in \eqref{eq:var2} while the gauge variation of $\alpha' c \,\Sigma_1 +\Omega_+^z \Sigma_0$ is zero, as we have already noted above in Section \ref{subsec:boundary_sf}.
Extremizing $\mathcal{S}$ with respect to $\Xi$ then yields back the action \eqref{eq:TheAction} subject to the linear relation \eqref{eq:lin_rel}.

\subsection{Open string}
\label{subsec:open}

Having obtained a general formula \eqref{eq:TheAction} for a gauge-invariant action at the massless level, let us now apply it in the case of the massless open-string fields in $d=26$ non-compact dimensions. Hence, we will consider the field content of the first line of \eqref{eq:masslessfields} for the bulk string field $\Psi$, namely
\begin{align}
     \Psi(\hat{x})=\Big[A_\mu(\hat{x}) \,c_1 \alpha_{-1}^\mu - i\sqrt{\tfrac{\alpha'}{2}} B(\hat{x}) \, c_0 \Big] |0\rangle\,.
\end{align}
Furthermore, we write the boundary string field in components as 
\begin{align}
     \Sigma_{k}(\hat{y})=\Big[-m^{(k)}_\mu(\hat{y}) \,c_1 \alpha_{-1}^\mu + i\sqrt{\tfrac{\alpha'}{2}} m^{(k)}(\hat{y}) \, c_0 \Big] |0\rangle
\end{align}
for $k=0,1$ where $m^{(k)}_\mu({y})$ and $m^{(k)}({y})$ are some boundary fields. Using \eqref{eq:openQ}, the gauge transformation laws for the bulk field and the boundary field \eqref{eq:bndy_gauge_tr} yield the following expression at the level of the target space fields
\begin{subequations}
\label{eq:gauge_open}
\begin{align}
    \delta_{\Lambda} A_{\mu} (x) &= \partial_{\mu}\lambda(x)\,,\\
    \delta_{\Lambda} B (x) &= \partial^2 \lambda(x)\,,\\
    \delta_{\Lambda} m^{(k)}_{\mu} (y) &= 0\,,\\
    \delta_{\Lambda} m^{(k)} (y) &= \p_z^{k+1} \lambda(x)|_{z=0}\,,
\end{align}
\end{subequations}
where we have written the gauge parameter as
\begin{align}
    \Lambda(\hat{x})=\frac{i}{\sqrt{2\alpha'}}\lambda(\hat{x})\ket{0}\,.
\end{align}
Before writing down the gauge-invariant action, we shall discuss the boundary field truncation of \eqref{eq:lin_rel}.
As explained in the previous sections, we are interested in the minimal extension of the starting action. Therefore, following this philosophy, we will be interested in as minimal a solution as possible of \eqref{eq:lin_rel} which, however, is still consistent with the gauge transformations \eqref{eq:gauge_open}. This leads us to setting
\begin{align}
    m^{(k)}_{\mu}(y)=0\,.
\end{align}
Substituting all ingredients, the gauge invariant action  \eqref{eq:TheAction} becomes 
\begin{align}
    S(A_\mu,B; m^{(0)},m^{(1)})&=\frac{\alpha'}{2}\int_M d^d x\,\bigg[\frac{1}{2}F_{\mu \nu}F^{\mu \nu}+\big(B-\partial_{\mu}A^{\mu}\big)^{2}\bigg]+\nonumber\\
    &\hspace{2cm}+ \alpha' \int_{\partial M}d^{d-1}y\,\big(A_z-m^{(0)}\big)\big(B-\partial_{a}A^a-m^{(1)}\big)\,.
\end{align}
We have rearranged the above expression to explicitly see the structure of the auxiliary fields: notice that we have completed the square for the Nakanishi-Lautrup field in the bulk term and factored out the expressions linear in $m^{(0)}$ and $m^{(1)}$ in the boundary term. Therefore, integrating out the auxiliary fields amounts to ignoring these terms. After doing so, one obtains the familiar gauge-invariant Maxwell action with a boundary
\begin{align}
    S^{\ast}(A_\mu)&=\frac{\alpha'}{4}\int_M d^d x\,F_{\mu \nu}F^{\mu \nu}.
\end{align}
Notice that we obtained a gauge-invariant action without additional degrees of freedom at the boundary. This means that in this case, we could have restored gauge invariance simply by adding a boundary term that depends only on the bulk fields, or, in other words, by considering only the cyclic action \eqref{eq:explicitS} with a different $\gamma^{\ast}$ (recalling that it is fixed up to BPZ-odd corrections).


\subsection{Closed string}\label{subsec:closed}

Let us now turn to discussing the construction of a gauge-invariant action for the massless closed string sector. Recalling \eqref{eq:masslessfields}, the level-$0$ part of the bulk closed-string field reads
\begin{align}
\Psi(\hat{x})&=\Big[-\frac{1}{2}e_{\mu\nu}(\hat{x}) \alpha_{-1}^\mu \bar{\alpha}_{-1}^\nu c_1\bar{c}_1+e(\hat{x})c_1c_{-1}+\bar{e}(\hat{x})\bar{c}_1\bar{c}_{-1}+\nonumber  \\&
    \hspace{3cm}+i\sqrt{\frac{\alpha'}{2}} c_0^+ \big( e_\mu(\hat{x}) c_1\alpha_{-1}^\mu+\bar{e}_\mu(\hat{x}) \bar{c}_1\bar{\alpha}_{-1}^\mu\big)\Big]|0\rangle\,.    
\end{align}
Following the notation of the previous section, we write the boundary string field as
\begin{align}
\Sigma_{k}(\hat{y})&=\Big[\frac{1}{2}m^{(k)}_{\mu\nu}(\hat{y}) \alpha_{-1}^\mu \bar{\alpha}_{-1}^\nu c_1\bar{c}_1-m^{(k)}(\hat{y})c_1c_{-1}-\bar{m}^{(k)}(\hat{y})\bar{c}_1\bar{c}_{-1}+\nonumber  \\&
    \hspace{3cm}-i\sqrt{\frac{\alpha'}{2}} c_0^+ \big( m^{(k)}_\mu(\hat{y}) c_1\alpha_{-1}^\mu+\bar{m}^{(k)}_\mu(\hat{y}) \bar{c}_1\bar{\alpha}_{-1}^\mu\big)\Big]|0\rangle 
\end{align}
for $k=0,1$, where $m^{(k)}_{\mu\nu}({y})$, $m^{(k)}({y})$, $\bar{m}^{(k)}({y})$, $m^{(k)}_\mu({y})$ and $\bar{m}^{(k)}_\mu(y)$ are some boundary fields.
Introducing the gauge parameter 
\begin{align}
    \Lambda(\hat{x})=\Big[ \frac{i}{\sqrt{2\alpha'}}\big(\xi_{\mu}(\hat{x})c_1\alpha^{\mu}_{-1}-\bar{\xi}_{\mu}(\hat{x})\bar{c}_{1}\bar{\alpha}^{\mu}_{-1}\big)-\mu(\hat{x})\Big]\ket{0}\,,
\end{align}
the gauge-transformation $\delta_{\Lambda}\Psi=\hat{Q}_\mathrm{B}\Lambda$ of the bulk string field translates into 
\begin{subequations}
\begin{align}
    \delta_{\Lambda} e_{\mu\nu}(x)&=\partial_{\mu} \bar{\xi}_{\nu}(x)+\partial_{\nu}\xi_{\mu}(x)\,,\\[1mm]
    \delta_{\Lambda} e_{\mu}(x)&=-\frac{1}{2}\partial^2\xi_\mu(x)-\partial_\mu\mu(x)\,,\\
    \delta_{\Lambda} \bar{e}_{\mu}(x)&=+\frac{1}{2}\partial^2\bar{\xi}_\mu(x)-\partial_\mu\mu(x)\,,\\
    \delta_{\Lambda}e(x)&= -\frac{1}{2}\partial_\mu \xi^\mu(x)-\mu(x)\,,\\
    \delta_{\Lambda}\bar{e}(x)&= \frac{1}{2}\partial_\mu \bar{\xi}^\mu(x)-\mu(x)\,,
\end{align}
\end{subequations}
for the bulk target fields, while the transformation laws \eqref{eq:bndy_gauge_tr} for the boundary string fields $\Sigma_0$, $\Sigma_1$ become
\begin{subequations}
\label{eq:gauge_closed_bndy}
\begin{align}
    \delta_{\Lambda} m_{\mu\nu}^{(k)}(y)&=0\,,\\[1mm]
    \delta_{\Lambda} m_{\mu}^{(k)}(y)&=-\frac{1}{2}\xi_{\mu}^{(k+1)}(y)\,,\\
    \delta_{\Lambda} \bar{m}_{\mu}^{(k)}(y)&=\frac{1}{2}\bar{\xi}_{\mu}^{(k+1)}(y)\,,\\
    \delta_{\Lambda} m^{(k)}(y)&=-\frac{1}{2}\xi^{(k)}_z(y)\,,\\
     \delta_{\Lambda} \bar{m}^{(k)}(y)&=\frac{1}{2}\bar{\xi}^{(k)}_z(y)\,,
\end{align}
\end{subequations}
Let us now impose linear relations between the boundary fields which are consistent with the condition \eqref{eq:lin_rel}.
Consistently with the gauge transformations listed in \eqref{eq:gauge_closed_bndy}, we will be setting
\begin{subequations}
    \begin{align}
        m_{\mu\nu}^{(k)}(y)&=0\,,\\
          m^{(k+1)}(y)-m_{z}^{(k)}(y)&=0\,,\\
        \bar{m}^{(k+1)}(y)-\bar{m}_{z}^{(k)}(y)&=0\,,
    \end{align}
\end{subequations}
for $k=0,1$. In particular, this allows us to get rid of both 
$m^{(1)}(y)$ and  $\bar{m}^{(1)}(y)$.
We are now ready to compute the gauge-invariant action, which, after a rather long series of manipulations, can be written as
\begin{align}
\label{eq:closedlevel0action}
&S(e_{\mu\nu},e_\mu,\bar{e}_\mu,e,\bar{e};m^{(0)},\bar{m}^{(0)},m^{(0)}_\mu,\bar{m}^{(0)}_\mu,m^{(1)}_\mu,\bar{m}^{(1)}_\mu)=\nonumber\\[1mm]
&\hspace{1cm}= \frac{\alpha'}{16}\int_{M}d^{d}x\, \Big[ -\frac{1}{4}\partial_{\rho}e_{\mu\nu}\partial^{\rho}e^{\mu\nu}+\frac{1}{4}\partial_{\rho}e_{\mu\nu}\partial^{\nu}e^{\mu \rho}+\frac{1}{4}\partial_{\rho}e_{\nu\mu}\partial^{\nu}e^{\rho\mu}\nonumber\\&\hspace{7cm}+\big(\partial_{\mu}(e-\bar{e})\big)^2-\frac{1}{2}\partial^\nu(e_{\nu\mu}+e_{\mu\nu})\partial^{\mu}(\bar{e}-e)\Big]\nonumber\\
    &\hspace{1cm}+ \frac{\alpha'}{16}\int_{\partial M}d^{d-1}y\, \Big[\frac{1}{2}(e_{az}+e_{za})\partial^{a}(\bar{e}-e-e_{zz})+\nonumber\\
    &\hspace{6cm}+\partial_{a}\big(m^{(0)}-\bar{m}^{(0)}\big)\partial_b\big((2\bar{e}-2e-e_{zz})\eta^{ab}-e^{ab}\big)\Big]+\nonumber\\[2mm]
    &\hspace{1cm}+S_\mathrm{aux}(e_{\mu\nu},e_\mu,\bar{e}_\mu,e,\bar{e};m^{(0)},\bar{m}^{(0)},m^{(0)}_\mu,\bar{m}^{(0)}_\mu,m^{(1)}_\mu,\bar{m}^{(1)}_\mu)\,,
\end{align}
where in $S_\mathrm{aux}$, we have collected all terms where the auxiliary fields contribute. These read
\begin{align}
    &S_\mathrm{aux}(e_{\mu\nu},e_\mu,\bar{e}_\mu,e,\bar{e};m^{(0)},\bar{m}^{(0)},m^{(0)}_\mu,\bar{m}^{(0)}_\mu,m^{(1)}_\mu,\bar{m}^{(1)}_\mu)=\nonumber\\[1.7mm]
    &\hspace{1cm}=\frac{\alpha'}{16}\int_{M}d^{d}x\, \Big[-\Big(e_{\mu}+\frac{1}{2}\partial^{\nu}e_{\mu\nu}-\partial_{\mu}\bar{e}\Big)^2-\Big(\bar{e}_{\mu}-\frac{1}{2}\partial^{\nu}e_{\nu\mu}-\partial_{\mu}e\Big)^2\Big]+\nonumber\\
    &\hspace{1cm}+ \frac{\alpha'}{16}\int_{\partial M}d^{d-1}y\, \Big[2\big(e_z-\bar{e}_z-m^{(1)}_z+\bar{m}^{(1)}_{z}+\partial^{2}_{a}(\bar{m}^{(0)}-m^{(0)})\big)\big(m_{z}^{(0)}-\bar{m}_{z}^{(0)}+\frac{1}{2}e_{zz}\big)\nonumber\\
&\hspace{2.9cm}+2\big(e^a-m^{(1)a}-\partial^a\bar{e}+\partial^a \bar{m}_{z}^{(0)}+\frac{1}{2}\partial_{b}e^{ab}\big)\big(m_{a}^{(0)}-\partial_a \bar{m}^{(0)}+\frac{1}{2}e_{az}\big)\nonumber\\[0.7mm]
&\hspace{2.9cm}+2\big(\bar{e}^a-\bar{m}^{(1)a}-\partial^a e+\partial^a m_{z}^{(0)}+\frac{1}{2}\partial_{b}e^{ba}\big)\big(\bar{m}_{a}^{(0)}-\partial_a 
m^{(0)}-\frac{1}{2}e_{za}\big)\Big]\,.\label{eq:Saux_closed}
\end{align}
Similarly to the open string case, in \eqref{eq:Saux_closed} we made the auxiliary field structure explicit by completing squares. In particular, in the bulk term of \eqref{eq:Saux_closed}, we can notice the presence of the auxiliary fields $e_\mu$ and $\bar{e}_\mu$. Integrating them out corresponds to setting the squares to zero. Moreover, in the second, third, and fourth line, we find the boundary auxiliary fields $m^{(0)}_{\mu}$ and $m^{(1)}_{\mu}$. All these terms vanish by solving the equations of motion for one of these fields.

To better understand the structure of \eqref{eq:closedlevel0action}, it is useful to explicitly write $e_{\mu \nu}$ as the sum of its symmetric and antisymmetric parts, namely
\begin{equation}
      e_{\mu \nu} = h_{\mu\nu}+b_{\mu \nu}\,,
\end{equation}
where we have introduced the symmetric part $h_{\mu\nu}=h_{\nu \mu}$ and the anti-symmetric part $b_{\mu\nu}=-b_{\nu \mu}$ of $e_{\mu\nu}$. Furthermore, let us introduce the fully antisymmetric tensor (3-form)
\begin{align}
    H_{\mu\nu\rho}=\p_\mu b_{\nu\rho}+\p_{\nu}b_{\rho \mu}+\p_{\rho}b_{\mu \nu}\,,
\end{align}
and perform the following field redefinition
\begin{subequations}
\begin{align}
    \frac{h}{2}&=\bar{e}-e\,,\\
    \frac{l}{2}&=\bar{m}^{(0)}-m^{(0)}\,.
\end{align}
\end{subequations}
These are the only combinations of $e,\bar{e}$ and $m^{(0)},\bar{m}^{(0)}$ entering the dynamical part of the action \eqref{eq:closedlevel0action}, the orthogonal combinations decouple. Correspondingly, it will be convenient to define rotated gauge parameters
\begin{subequations}
    \begin{align}
      \lambda_\mu&=\frac{1}{2}(\bar{\xi}_\mu-\xi_{\mu})\,,\\
      \tau_\mu&=\frac{1}{2}(\bar{\xi}_\mu+\xi_{\mu})\,.
    \end{align}
\end{subequations}
Using these definitions and integrating out the auxiliary fields, we obtain the action
\begin{align}\label{eq:level0closedactionfinal}
    &S^{\ast}\left(h_{\mu \nu},b_{\mu \nu},h;l\right)=\nonumber\\
    &\hspace{0.4cm}=-\frac{\alpha^\prime}{16}\int_{M}d^d x\, \bigg(\frac{1}{12}H_{\mu\nu\rho}H^{\mu\nu\rho}\bigg)+\nonumber\\
    &\hspace{1cm}-\frac{\alpha^\prime}{16}\int_{M}d^d x\,\bigg(\frac{1}{4}\partial_\rho h_{\mu\nu}\partial^\rho h^{\mu\nu}-\frac{1}{4}\partial_{\mu }h\partial^{\mu }h-\frac{1}{2}\partial_{\rho}h_{\mu \nu}\partial^\nu h^{\mu\rho}+\frac{1}{2}\partial^{\mu}h_{\mu \nu}\partial^\nu h\bigg)+\nonumber\\
    &\hspace{1cm}+\frac{\alpha^\prime}{16}\int_{\partial M}d^{d-1} y\, \bigg[\frac{1}{2}h_{az}\partial^{a}\big(h-2h^{zz}\big)+\frac{1}{2}\left(\left(h-h_{zz}\right)\eta^{ab}-h^{ab}\right)\partial_{a}\partial_{b}l\bigg],
\end{align}
which is invariant under the gauge transformations
\begin{subequations}
    \begin{align}
        \delta_{\Lambda}b_{\mu\nu}(x)&=\partial_{\mu}\lambda_{\nu}(x)-\partial_{\nu}\lambda_{\mu}(x)\,,\\
        \delta_{\Lambda}h_{\mu\nu}(x)&=\partial_{\mu}\tau_{\nu}(x)+\partial_{\nu}\tau_{\mu}(x)\,,\label{eq:deltah}\\
        \delta_{\Lambda}h(x)&=2\partial_{\mu}\tau^\mu(x)\,,\\
        \delta_{\Lambda}l(y)&=2\tau_{z}(y)\,.\label{eq:deltal}
    \end{align}
\end{subequations}
Finally, by introducing the physical gauge-invariant dilaton
\begin{align}
    \phi= \frac{1}{4}\left(\tensor{h}{^{\mu}_{\mu}}-h\right),
\end{align}
this action can be rewritten as
\begin{align}\label{eq:level0closedactionfinal2}
    &S^{\ast}\left(h_{\mu \nu},b_{\mu \nu},h(\phi);l\right)=\nonumber\\
    &\hspace{0.4cm}=-\frac{\alpha^\prime}{16}\int_{M}d^d x\, \bigg(\frac{1}{12}H_{\mu\nu\rho}H^{\mu\nu\rho}\bigg)+\nonumber\\
    &\hspace{1cm}-\frac{\alpha^\prime}{16}\int_{M}d^d x\,\bigg(\frac{1}{4}\partial_\rho h_{\mu\nu}\partial^\rho h^{\mu\nu}-\frac{1}{2}\partial_{\rho}h_{\mu \nu}\partial^\nu h^{\mu\rho}+\frac{1}{4}\p^\mu\big(2h_{\mu \nu}-\eta_{\mu \nu}\tensor{h}{^{\rho}_{\rho}}\big)\p^\nu \tensor{h}{^{\lambda}_{\lambda}}\nonumber\\
    &\hspace{7cm}-2\p^\mu\big(h_{\mu\nu}-\eta_{\mu \nu}\tensor{h}{^{\rho}_{\rho}}\big)\p^{\nu}\phi-4\p_\mu\phi\p^\mu\phi\bigg)+\nonumber\\
    &\hspace{1cm}+\frac{\alpha^\prime}{16}\int_{\partial M}d^{d-1} y\, \bigg[\frac{1}{2}h_{az}\partial^{a}\big(\tensor{h}{^{b}_{b}}-h_{zz}-4\phi\big)+\frac{1}{2}\left(\left(\tensor{h}{^{c}_{c}}-4\phi\right)\eta^{ab}-h^{ab}\right)\partial_{a}\partial_{b}l\bigg].
\end{align}
Looking at \eqref{eq:level0closedactionfinal2}, we can recognize the standard linearized low-energy effective string action for the graviton, Kalb-Ramond, and dilaton fields in the bulk, with a boundary term involving an additional boundary degree of freedom $l$ which only couples to the graviton-dilaton sector.

We will see in Appendix \ref{app:linear} that upon setting to zero the dilaton and the Kalb-Ramond field, the action \eqref{eq:level0closedactionfinal2} coincides with the linearization of the Einstein-Hilbert action supplemented with the Gibbons-Hawking-York term, which is expanded around flat space with a flat boundary. In particular, representing the shape of a fluctuating boundary by a function $z(y)$ (which, in the absence of fluctuations, becomes simply $z=0$), we will see that setting
\begin{align}
z(y)=-\frac12 l(y)\,,\label{eq:zl}
\end{align}
as in \eqref{eq:fl}, the EH+GHY action expanded to second order in $h_{\mu\nu}$ and $l$ precisely reproduces the pure-graviton part of \eqref{eq:level0closedactionfinal2}. One can also check that defining $l$ as in \eqref{eq:zl}, the gauge transformation law \eqref{eq:deltal} follows from the same infinitesimal diffeomorphism which gives rise to the gauge transformation \eqref{eq:deltah} of the metric fluctuation. We can therefore conclude that the boundary field $l(y)$ can be associated with  the transverse displacement of the boundary. 
\section{The tensionless limit}\label{sec:tensionless}

In this section we will demonstrate that the algebraic structure, which we identified in Section~\ref{sec:massless} for the case of massless excitations of both the open and the closed string, is also capable of accommodating the tensionless limit ($\alpha'\to\infty$) of the full string 
field theory. 
It is well-known (see for example \cite{Fronsdal:1978rb, Francia:2002pt, Barnich:2005ga} and reference therein) that, at least at the free level, the tensionless limit can be used to describe free massless higher-spin theories. This contains the case of the massless spin-2, so this is an infinite extension of free gravity, which is however different from the standard string theory as the higher spin states are still massless \footnote{This can be understood as an unbroken-phase of string theory, with $\alpha'$ playing the role of the symmetry-breaking order parameter \cite{Gross:1987ar}.}.
As a result, we will be able to write down an action for the full tensionless SFT on a flat target with a flat boundary. As an application, we will work out an explicit form of the free action for the leading Regge trajectory of OSFT in the limit $\alpha'\to \infty$.

\subsection{Tensionless BRST operator}\label{subsec:tensionless_brst}
The Hilbert space of the tensionless string is the same as that of the tensile string and the same is true for the oscillator algebra. 
The big difference is in the algebra of constraints which, in the tensionless case, is a contraction of the matter Virasoro algebra
\begin{subequations}
\begin{align}
L_0\quad&\longrightarrow\quad \lim_{\alpha'\to\infty}\frac1{\alpha'}L_0=\frac1{2\alpha'}\alpha_0^2\equiv l_0\,,\\
L_n\quad&\longrightarrow\quad \lim_{\alpha'\to\infty}\frac1{\sqrt{\alpha'}}L_n=\frac1{\sqrt{\alpha'}}\alpha_0\cdot\alpha_n\equiv l_n\,,
\end{align}
\end{subequations}
where, as usual,
\begin{subequations}
\begin{align}
\alpha_0^\mu&=\sqrt{2\alpha'} p^\mu\,, \hspace{3cm} \textrm{(open strings)}\\
\alpha_0^\mu&=\bar\alpha_0^\mu=\sqrt{\frac{\alpha'}{2}} p^\mu\,. \hspace{2cm} \textrm{(closed strings)}
\end{align}
\end{subequations}
What remains of the Virasoro algebra of constraints is readily found to be
\begin{align}
[l_n,l_m]=\gamma_{nm}^k l_k =   2n\,l_0\,\delta_{n+m}\,,
\end{align}
where the structure constants are $\gamma_{nm}^k=2n\delta_{n+m}\delta_{k0}$. Notice in particular the absence of the central extension. The corresponding holomorphic BRST charge is constructed standardly as follows 
\begin{align}
Q&=\sum_n c_{-n} l_{n}+\frac12\sum_n  c_{-n}l^{\rm gh}_n\,,
\end{align}
where the ghost modes $l^{\rm gh}_n$ are defined as
\begin{align}
    l^{\rm gh}_n&=-\gamma_{nm}^k\,c_{-m} b_k=-2nc_nb_0\,.
\end{align}
For the closed string there is an isomorphic anti-holomorphic algebra, leading to the final expressions
\begin{align}
Q_{\rm open}&= c_0p^2+\sqrt2 p_\mu\sum_{n\neq 0}c_{-n}\alpha_n^\mu-2b_0\sum_{n>0}nc_{-n}c_n\,,\label{eq:Q_op_tl}
\end{align}
for the open-string BRST operator and
\begin{align}
Q_{\rm closed}
&=c_0^+\frac{p^2}{2} + \frac1{\sqrt2}p_{\mu}\sum_{n\neq 0}\left(c_{-n}\alpha_n^\mu+\bar c_{-n}\bar\alpha_n^\mu\right)-b_0^+
\sum_{n>0}n\left(c_{-n}c_n+\bar c_{-n}\bar c_n\right)\,,\label{eq:Q_cl_tl}
\end{align}
for the closed-string BRST operator. Note that in writing down \eqref{eq:Q_cl_tl}, we have recalled that the closed-string Hilbert space is level matched and, in particular, that $b_0=\bar b_0$.
It is easy to show that nilpotency doesn't require any critical dimension in either open or closed case.

\subsection{Gauge-invariant action}

It follows from the discussion in Section \ref{subsec:tensionless_brst} that the BRST operator of the tensionless string field theory on $M\subset \mathbb{R}^d$ can be decomposed as
\begin{align}
    Q_\mathrm{B} =- c \p_z^2-\Omega^z \p_z +\tilde{Q}\,.
\end{align}
As in Section \ref{sec:massless}, we are using a cartesian coordinate system $x^\mu =(y^a,z)$ for $\mathbb{R}^d$ where the directions $y^a$ are extending along the flat boundary $\p M$ which is located at $z=0$. In particular, we recognize
\begin{align}
\label{eq:openQtensionless}
    c &= c_0\,,\nonumber\\[2.4mm]
    \Omega^\mu &= i\sqrt{2}\sum_{n\neq 0}c_{-n} \alpha_{n}^\mu \,,\hspace{5.8cm} \text{(open string)}\\[-1.2mm]
    \tilde{Q} &=- c \p_a\p^a-\Omega^a \p_a-2b_0 \sum_{n>0} nc_{-n}c_{n}\,,\nonumber
\end{align}
for the open string, as well as
\begin{align}
\label{eq:closedQtensionless}
    c &= \frac{1}{2}c_0^+\,,\nonumber\\
    \Omega^\mu &= \frac{i}{\sqrt{2}}\sum_{n\neq 0}\big(c_{-n} \alpha_{n}^\mu+\bar{c}_{-n} \bar{\alpha}_{n}^\mu\big) \,,\hspace{3.8cm} \text{(closed string)}\\
    \tilde{Q}&=- c \p_a\p^a-\Omega^a \p_a-b_0^+ \sum_{n>0}n\big(c_{-n}c_{n}+\bar{c}_{-n}\bar{c}_n\big)\,.\nonumber
\end{align}
for the closed string. Furthermore, we notice that in both \eqref{eq:openQtensionless} and \eqref{eq:closedQtensionless}, one can split $\Omega^\mu$ into $\Omega^\mu_+$ and $\Omega^\mu_-$ as in \eqref{eq:Omegasplit} (where we recall that $\Omega^\mu_+$ contains only the positively-moded matter oscillators and vice versa for $\Omega^\mu_-$) such that both $\Omega^z_\pm$ separately anti-commute with $\tilde{Q}$. As we have seen in Section \ref{sec:massless}, this is enough to guarantee that the action
\begin{align}
    & S(\Psi;\Sigma_0,\Sigma_1)=\nonumber\\
    &\hspace{0.8cm}=\frac{1}{2}\int_{M}  d^d x \,\omega'\big(\p_\mu\Psi(x), c\p^\mu\Psi(x)\big)+\frac{1}{2} \int_{M}  d^d x\, \omega'\big(\Psi(x),(Q'- 2\Omega_-^\mu\p_\mu)\Psi(x)\big)+\nonumber\\
    &\hspace{5.4cm}+\int_{\p M}  d^{d-1} y \,\omega'\big(\Psi(y)+\Sigma_1(y), \tilde{Q}\Sigma_0(y)-\Omega_-^z \Sigma_1(y)\big)\,,\label{eq:TheActionTL}
\end{align}
is invariant under the gauge transformation\footnote{Recall that we define $\Lambda^{(k)}(y)\equiv \p_z^k\Lambda(x)|_{z=0}$.}
\begin{subequations}
\label{eq:bndy_gauge_tr_TL}
    \begin{align}
        \delta_\Lambda \Psi(x)&= Q_\mathrm{B}\Lambda(x)\,,\\[1mm]
        \delta_\Lambda \Sigma_0(y) &=  c \Lambda^{(1)}(y)+\Omega_+^z\Lambda^{(0)}(y)\,,\\[1.0mm]
        \delta_\Lambda \Sigma_1(y) &=  c \Lambda^{(2)}(y)+\Omega_+^z\Lambda^{(1)}(y)\,,
    \end{align}
\end{subequations}
assuming the linear relation
\begin{align}
    \alpha' c \,\Sigma_1(y) +\Omega_+^z \Sigma_0(y)=0\label{eq:lin_rel_TL}
\end{align}
between the boundary string fields $\Sigma_0(y)$ and $\Sigma_1(y)$.

\subsection{Example: Leading Regge trajectory of tensionless OSFT}
In this section, we will see the action \eqref{eq:TheActionTL} at work for the explicit example of the leading Regge trajectory of OSFT in the tensionless limit. That is to say, we will be considering only states generated by the oscillators $\alpha_{-1}^{\mu}$ so that at each level $s-1$ (with $s>1$), we will consider the so-called higher-spin triplet \cite{Francia:2002pt}
\begin{equation}
    \varphi_{\mu_1\ldots\mu_s}(x)\,,\qquad C_{\mu_1\ldots\mu_{s-1}}(x)\,,\qquad D_{\mu_1\ldots\mu_{s-2}}(x)\,.\label{eq:HStriplet}
\end{equation}
This is a consistent sub-sector characterized by three symmetric tensor fields of rank $s$, $s-1$ and $s-2$, respectively, which can be collected in the level $s-1$ string field as
\begin{align}\label{eq:HStripletbulk}
    \Psi_{s-1}(\hat{x})&=\Big[\frac{1}{2}  \varphi_{\mu_1\ldots\mu_s}(\hat{x})\alpha_{-1}^{\mu_1}\ldots\alpha_{-1}^{\mu_s}c_1+\frac{i}{\sqrt{2}}C_{\mu_1\ldots\mu_{s-1}}(\hat{x})\alpha_{-1}^{\mu_1}\ldots\alpha_{-1}^{\mu_{s-1}}c_0\nonumber\\&\hspace{6cm}+\frac{1}{2}D_{\mu_1\ldots\mu_{s-2}}(\hat{x})\alpha_{-1}^{\mu_1}\ldots\alpha_{-1}^{\mu_{s-2}}c_{-1}\Big]\ket{0}\,.
\end{align}
Note that this approach can be trivially extended for $s=1$ by setting $D=0$ and identifying $\varphi_\mu$ as the gauge field $A_\mu$ and $C$ as (minus) the Nakanishi-Lautrup field $B$.  

As for the boundary degrees of freedom, we introduce two boundary higher-spin triplets
\begin{equation}
    m^{(k)}_{\mu_1\ldots\mu_s}(y)\,,\qquad N^{(k)}_{\mu_1\ldots\mu_{s-1}}(y)\,,\qquad M^{(k)}_{\mu_1\ldots\mu_{s-2}}(y)\,,
\end{equation}
for $k=0, 1$, which are totally symmetric rank $s$, $s-1$ and $s-2$ tensor fields. Therefore, the boundary string field has the following structure
\begin{align}\label{eq:HStripletbndy}
    \Sigma_{k,\,s-1}(\hat{y})&=\Big[-\frac{1}{2}  m^{(k)}_{\mu_1\ldots\mu_s}(\hat{y})\alpha_{-1}^{\mu_1}\ldots\alpha_{-1}^{\mu_s}c_1-\frac{i}{\sqrt{2}}N^{(k)}_{\mu_1\ldots\mu_{s-1}}(\hat{y})\alpha_{-1}^{\mu_1}\ldots\alpha_{-1}^{\mu_{s-1}}c_0+\nonumber\\&\hspace{6cm}-\frac{1}{2}M^{(k)}_{\mu_1\ldots\mu_{s-2}}(\hat{y})\alpha_{-1}^{\mu_1}\ldots\alpha_{-1}^{\mu_{s-2}}c_{-1}\Big]\ket{0}\,.
\end{align}
Given these definitions, the gauge-transformation laws \eqref{eq:bndy_gauge_tr_TL} applied to the gauge parameter
\begin{align}
    \Lambda_{s-1}(\hat{x})=\frac{i}{\sqrt{2}}\lambda_{\mu_1\ldots\mu_{s-1}}(\hat{x})\alpha_{-1}^{\mu_1}\ldots\alpha_{-1}^{\mu_{s-1}}\ket{0}\,,
\end{align}
yield
\begin{subequations}
    \begin{align}
        \delta_{\Lambda} \varphi_{\mu_1\ldots\mu_{s}}(x)&=2\p_{(\mu_1}\lambda_{\mu_2\ldots\mu_{s})}(x)\,,\\
        \delta_{\Lambda} C_{\mu_1\ldots\mu_{s-1}}(x)&=-\p^2\lambda_{\mu_1\ldots\mu_{s-1}}(x)\,,\\
        \delta_{\Lambda} D_{\mu_1\ldots\mu_{s-2}}(x)&=2(s-1)\p^{\mu}\lambda_{\mu\mu_1\ldots\mu_{s-2}}(x)\,,\\
        \delta_{\Lambda}m^{(k)}_{\mu_1\ldots\mu_{s}}(y)&=0\,,\\
        \delta_{\Lambda} N^{(k)}_{\mu_1\ldots\mu_{s-1}}(y)&=-\p^{k+1}_{z} \lambda_{\mu_1\ldots\mu_{s-1}}(x)\vert_{z=0}\,,\\
        \delta_{\Lambda} M^{(k)}_{\mu_1\ldots\mu_{s-2}}(y)&=2(s-1)\p^{k}_{z}\lambda_{z\mu_1\ldots\mu_{s-2}}(x)\vert_{z=0}\,,
    \end{align}
\end{subequations}
at the level of the target space fields. We use the notation $(\mu_1 \ldots \mu_{s})$ to denote the symmetrization over the indices $\mu_1, \ldots ,\mu_{s}$ weighted by $(s!)^{-1}$. Consistent with the level-0 examples discussed in Sections \ref{subsec:closed} and \ref{subsec:open} above, we will be also imposing the gauge-invariant truncations 
\begin{subequations}\label{eq:truncationHStriplet}
    \begin{align}
       m^{(k)}_{\mu_1\ldots\mu_{s}}(y)&=0\,,\\ 
       M^{(k+1)}_{\mu_1\ldots\mu_{s-2}}(y)+2(s-1)N^{(k)}_{z\mu_1\ldots\mu_{s-2}}(y)&=0\,,
    \end{align}
\end{subequations}
of the boundary fields for $k=0,1$ which, in particular, ensure the validity of the linear relation \eqref{eq:lin_rel_TL}. 

Given these ingredients, by substituting the string fields \eqref{eq:HStripletbulk} and \eqref{eq:HStripletbndy} in \eqref{eq:bndy_gauge_tr_TL} into the action \eqref{eq:TheActionTL} and imposing the condition \eqref{eq:truncationHStriplet}, we obtain a gauge-invariant action
\begin{align}
\label{eq:HStripletaction}
&S( \varphi, C, D;  N^{(0)},M^{(0)},N^{(1)})=\nonumber\\
&\hspace{0.4cm}= \frac{s!}{2}\int_{M}d^{d}x\, \Big[\frac{1}{4}\p_\mu\varphi_{\mu_1\ldots\mu_{s}}\p^\mu\varphi^{\mu_1\ldots\mu_{s}}-\frac{s}{4}\p_\nu\varphi_{\mu\mu_1\ldots\mu_{s-1}}\p^\mu\varphi^{\nu\mu_1\ldots\mu_{s-1}}\nonumber\\
&\hspace{3cm}-\frac{1}{4(s-1)}\p_\mu D_{\mu_1\ldots\mu_{s-2}}\p^\mu D^{\mu_1\ldots\mu_{s-2}}+\frac{1}{2}\p^\mu\varphi_{\mu \mu_1\ldots\mu_{s-1}}\p^{\mu_1}D^{\mu_2\ldots\mu_{s-1}}\Big]\nonumber
\\[1mm]
    &\hspace{1cm}+\frac{s!}{2}\int_{\partial M}d^{d-1}y\sum_{n=0}^{s-1}\binom{s-1}{n}\, \Big[\varphi_{z\ldots za_{1}\ldots a_{n}a}\p^a\Big(\frac{s}{2}\varphi^{z\ldots za_{1}\ldots a_{n}}+E\big(\varphi,M^{(0)}\big)^{z\ldots za_{1}\ldots a_{n}}\Big)\nonumber\\
    &\hspace{3.0cm}-\frac{s-n-1}{2s(s-1)^2}\Big(D^{z\ldots za_{1}\ldots a_{n}}+2(s-1)E\big(\varphi,M^{(0)}\big)^{z\ldots za_{1}\ldots a_{n}}\Big)\times\nonumber \\
    &\hspace{5cm}\times\p^a\Big(\p_a M^{(0)}_{z\ldots za_{1}\ldots a_{n}}+2(s-1)E\big(\varphi,M^{(0)}\big)_{z\ldots za_{1}\ldots a_{n}a}\Big)\Big]\nonumber\\[2mm]
    &\hspace{1cm}+S_\mathrm{aux}( \varphi, C, D;  N^{(0)},M^{(0)}, N^{(1)})
\end{align}
for the higher-spin triplet \eqref{eq:HStriplet} on the half-space $M\subset \mathbb{R}^d$ which is bounded by a codimension-one hyperplane $\p M$.
Similarly to the examples at level 0 presented above, we have completed the squares for all auxiliary fields and collected them in $S_{\rm aux}$ which then vanishes upon solving the corresponding equations of motion. These read
\begin{equation}
    s\p^\mu \varphi_{\mu \mu_1\ldots\mu_{s-1}}-\p_{(\mu_{1}}D_{\mu_2\ldots\mu_{s-1})}+2C_{\mu_1\ldots\mu_{s-1}}=0\,,
\end{equation}
for the bulk auxiliary field $C$, and 
\begin{align}
\label{eq:bndy_aux}
       &N^{(0)}_{z\ldots za_{1}\ldots a_{n}}=\nonumber\\
    &\hspace{1cm}=E(\varphi,M^{(0)})_{z\ldots za_{1}\ldots a_{n}}\coloneqq\frac{s}{2(s-n)}\left(\frac{n}{s(s-1)}\p_{(a_{1}}^{\,}M_{a_{2}\ldots a_{n})z\ldots z}^{(0)}-\varphi_{z\ldots za_{1}\ldots a_{n}}\right)
\end{align}
for the boundary auxiliary field $N^{(0)}$. Looking at \eqref{eq:bndy_aux}, we can notice that the on-shell value of the field $N^{(0)}$ is given by different expressions depending on the number of indices in the direction $z$ perpendicular to the boundary. This motivates our choice to split the boundary contribution in \eqref{eq:HStripletaction} as the sum of terms with a fixed number of indices in the $z$ direction.

Let us now specialize to the level $1$ open string sector ($s=2$). In this case, the gauge-invariant action \eqref{eq:HStripletaction} reads
\begin{align}
\label{eq:HStripletactionlevel1}
&S^{(s=2)}( \varphi, C, D;  N^{(0)},M^{(0)},N^{(1)})=\nonumber\\[1mm]
&\hspace{0.4cm}= -\frac{1}{2}\int_{M}d^{d}x\, \Big[-\frac{1}{2}\p_\rho\varphi_{\mu\nu}\p^\rho\varphi^{\mu\nu}+\p_\nu\varphi_{\mu\rho}\p^\mu\varphi^{\nu\rho}+\frac{1}{2}\p_\mu D\p^\mu D-\p^\mu \varphi_{\mu\nu}\p^\nu D\Big]\nonumber\\
    &\hspace{2cm}+\frac{1}{2}\int_{\partial M}d^{d-1}y\, \Big[\varphi_{za}\p^a(2\varphi_{zz}-D)-\big((D-\varphi_{zz})\eta_{ab}-\varphi_{ab}\big)\p^a\p^b M^{(0)}\Big]\nonumber\\[2mm]
    &\hspace{2cm}+S_\mathrm{aux}( \varphi, C, D;  N^{(0)},M^{(0)}, N^{(1)})\,,
\end{align}
where we have collected all contributions from auxiliary fields into
\begin{align}
    &S^{(s=2)}_{\rm aux}( \varphi, C, D;  N^{(0)},M^{(0)},N^{(1)})=\nonumber\\[1mm]
    &\hspace{0.4cm}=\frac{1}{4}\int_{M}d^{d}x\, \Big(\p^\mu \varphi_{\mu\nu}-\frac{1}{2}\p_\mu D+C_{\mu}\Big)^2\nonumber\\
    &\hspace{1cm}+ \int_{\partial M}d^{d-1}y\, \Big[\big(N^{(1)a}-C^{a}+\frac{1}{2}D+N^{(0)}_{z}-\p_{b}\varphi^{ba}\big)\big(N^{(0)}_a+\varphi_{za}-\frac{1}{2}\p_a M^{(0)}\big)\nonumber\\
&\hspace{6cm}+\big(N^{(1)z}-C^{z}-\frac{1}{2}\p_{a}^2 M^{(0)}\big)\big(2N^{(0)}_z+\varphi_{zz}\big)
\Big]\,.\label{eq:Saux_HStripletlevel1}
\end{align}
Therefore, by classically integrating out the auxiliary fields (i.e.\ setting $S_{\rm aux} = 0$) and renaming the remaining fields as follows
\begin{subequations}
    \begin{align}
    4 \varphi^{\mu \nu}(x) &=h^{\mu \nu}(x)\,,\\
     4D(x) &= h(x)\,,\\
    4 M^{(0)}(y)&=l(y)\,,
    \end{align}
\end{subequations}
we recover the already-familiar action \eqref{eq:closedlevel0action} describing the dynamics of the graviton and the dilaton, which we have already discussed in the context of the massless sector in closed SFT.





\section{Discussion and outlook for the full SFT}\label{sec:disc}
In this paper we have obtained a gauge-invariant construction of boundary terms for the massless level, as well as for the tensionless limit of both the open and closed string field theory. Our boundary terms generally involve additional boundary string fields whose gauge transformation properties compensate for the failure of the bulk action to be gauge invariant by itself. Similarly to the example of the abelian Chern-Simons theory which was reviewed in detail in Section \ref{sec:1}, it would be interesting to further investigate the dynamics of the excitations (edge modes) encoded in these boundary string fields.

We are now faced with addressing the remaining infinite massive spectrum of the full string theory. Since the problem of giving a well-defined variational principle has been solved for the full string theory in Section \ref{sec:position} (see also \cite{atakan,georg}), it remains to show if (and how) one can add appropriate boundary terms which would make the bulk-boundary theory fully gauge-invariant. As in the massless and tensionless case, one should expect that these should be generally involving additional boundary string fields. However, it should be pointed out that, at least at the level of free theory, the gauge symmetry of the massive levels does not straightforwardly translate into a standard space-time gauge symmetry: the physical states are represented by target fields which now pick up mass terms and these generally violate gauge invariance. Moreover this massive gauge symmetry is only realized in the critical dimension $d=26$: starting from level 1, the BRST charge is only nilpotent for a vanishing total worldsheet central charge. At the same time, the algebraic structure entering the full BRST charge is generally more involved than for the massless and tensionless truncations which were discussed in this paper. As a consequence, the setting presented in this paper does not fully apply anymore. In particular, the definition \eqref{eq:Gamma_star_plus} of $\Gamma^*$ that we used in the massless sector, as well as its all-level generalization utilized in Section \ref{sec:tensionless}, fail to anti-commute with the full BRST charge, namely
\begin{align}
[\Gamma^*,Q_\mathrm{B}]\neq 0\,. \quad\textrm{(massive levels)}
\end{align}
Nevertheless, there appear to be several possibilities on how to overcome this obstacle. We will report on these in \cite{future}, but
to give an anticipation of our results, let us write down a gauge-invariant action for the first massive level of OSFT on a flat half-space $M\subset \mathbb{R}^d$ with a flat boundary $\p M$. This turns out to read
\begin{align}
&S_\text{lev\,1}( \varphi_{\mu\nu}, C_\mu,D,a_\mu,a; M^{(0)})=\nonumber\\[1mm]
&\hspace{0.4cm}= \frac{\alpha^\prime}{2}\int_{M}d^{d}x\, \bigg\{\tfrac{1}{2}\p_\rho\varphi_{\mu\nu}\p^\rho\varphi^{\mu\nu}+\tfrac{1}{2}\tfrac{1}{\alpha^\prime}\varphi^{\mu\nu}\varphi_{\mu\nu}-\tfrac{1}{2}\p_\mu D\p^\mu D-\tfrac{1}{2}\tfrac{1}{\alpha^\prime} D^2\nonumber\\
&\hspace{9cm}+\p^\mu \varphi_{\mu\nu}\p^\nu D-\p_\nu\varphi_{\mu\rho}\p^\mu\varphi^{\nu\rho}+\nonumber\\
&\hspace{3cm}+\tfrac{1}{4}(\p^\mu a^\nu-\p^\nu a^\mu)(\p_\mu a_\nu-\p_\nu a_\mu)+\nonumber\\
&\hspace{3cm}+\sqrt{\tfrac{1}{2\alpha^\prime}}\Big[2\eta^{\rho\sigma}(\p_\rho \varphi_{\mu\sigma}-\p_\mu \varphi_{\rho\sigma})-\tfrac{5}{2}\p_\mu(D-\eta_{\rho\sigma}\varphi^{\rho\sigma})\Big]a^\mu+\nonumber\\
&\hspace{3cm}
-\tfrac{1}{16}\tfrac{1}{\alpha^\prime}(5D-\eta^{\mu\nu}\varphi_{\mu\nu})(D-\eta^{\mu\nu}\varphi_{\mu\nu})\bigg\}\nonumber\\
    &\hspace{1cm}+\frac{\alpha^\prime}{2}\int_{\partial M}d^{d-1}y\, \bigg\{\Big(2\p^a\varphi_{zz}-\p^aD-2a^a\sqrt{\tfrac{1}{2\alpha^\prime}}\Big)\varphi_{za}+\nonumber\\
    &\hspace{3cm}-\Big[\p^a\p_a(D-\varphi_{zz})-\p^a\p^b\varphi_{ab}+\nonumber\\
    &\hspace{5cm}+\p_a a^a \sqrt{\tfrac{2}{\alpha^\prime}}+\tfrac{1}{4}\tfrac{1}{\alpha^\prime}\eta_{ab}\varphi^{ab}-\tfrac{5}{4}\tfrac{1}{\alpha^\prime}D+\tfrac{5}{4}\tfrac{1}{\alpha^\prime}\varphi_{zz}\Big] M^{(0)}\bigg\}+\nonumber\\
    &\hspace{1cm}+S_\text{aux}( \varphi_{\mu\nu}, C_\mu,D,a_\mu,a; M^{(0)})\,.\label{eq:massive_lev1}
\end{align}
Here $\varphi_{\mu\nu}, C_\mu, D$ are the fields of the spin-2 triplet of target fields (which was already encountered in Section \ref{sec:tensionless} in the limit $\alpha^\prime\to \infty$) which correspond to the states $c_1\alpha_{-1}^\mu\alpha^\nu_{-1}|0\rangle$, $c_0\alpha_{-1}^\mu|0\rangle$, $c_{-1}|0\rangle$, while $a_\mu, a$ arise as expansion coefficients corresponding to the level-1 states $c_1\alpha^\mu_{-2}|0\rangle$ and $b_{-2}c_1 c_0|0\rangle$. All dependence on the auxiliary fields $C_\mu$ and $a$ is packaged inside
\begin{align}
   & S_\text{aux}( \varphi_{\mu\nu}, C_\mu,D,a_\mu,a; M^{(0)})=\nonumber\\
    &\hspace{1cm}=\frac{\alpha^\prime}{2}\int_{M}d^{d}x\,\bigg[\Big(C^\nu + \p_\mu\varphi^{\mu\nu}-\tfrac{1}{2}\p^\nu D-\sqrt{\tfrac{1}{2\alpha^\prime}}a^\nu\Big)^2+\nonumber\\
    &\hspace{6.0cm}+2\Big(a-\tfrac{1}{2}\p_\mu a^\mu -\tfrac{1}{4}\eta^{\mu\nu}\varphi_{\mu\nu}\sqrt{\tfrac{1}{2\alpha^\prime}}+\tfrac{3}{4}D\sqrt{\tfrac{1}{2\alpha^\prime}}\Big)^2\bigg]\,,
\end{align}
which can be dropped upon substituting for the on-shell values of $C_\mu$ and $a$. In $d=26$ dimensions, the action \eqref{eq:massive_lev1} is fully invariant under the gauge transformation
\begin{subequations}
\label{eq:massive_gauge_bulk}
\begin{align}
\delta_\xi \varphi_{\mu\nu} &= \p_\mu \xi_\nu+\p_\nu\xi_\mu - \tfrac{1}{2}\eta_{\mu\nu}\sqrt{\tfrac{2}{\alpha^\prime}}\xi\,,\\
\delta_\xi C_\mu &=-\Big(\p^\mu\p_\mu-\tfrac{1}{\alpha^\prime}\Big)\xi_\mu\,,\\
\delta_\xi D &=2\p_\mu \xi^\mu - 3\sqrt{\tfrac{2}{\alpha^\prime}}\xi\,,\\
\delta_\xi a_\mu &=2\p_\mu\xi+\sqrt{\tfrac{2}{\alpha^\prime}}\xi_\mu\,,\\[0.4mm]
\delta_\xi a &=\Big(\p^\mu\p_\mu-\tfrac{1}{\alpha^\prime}\Big)\xi\,,
\end{align}
\end{subequations}
as well as
\begin{align}
    \delta_\xi M^{(0)}(y) &= 2\xi_z(y)\,.\label{eq:gauge_massive_M}
\end{align}
The bulk part \eqref{eq:massive_gauge_bulk} of the gauge transformation can be generated by acting with the full BRST charge $Q_\mathrm{B}$ of OSFT on the gauge-parameter string field
\begin{align}
    \Lambda(x)=i\sqrt{\tfrac{1}{2\alpha^\prime}}\xi_\mu(x)\alpha_{-1}^\mu|0\rangle-\sqrt{\tfrac{1}{2\alpha^\prime}}\xi(x) b_{-2}c_1|0\rangle\,.
\end{align}
It is easy to check that by performing the tensionless limit, that is sending $\alpha'\to \infty$, the action \eqref{eq:massive_lev1}, as well as the gauge transformations \eqref{eq:massive_gauge_bulk} and \eqref{eq:gauge_massive_M} reduce to the expected massless spin-two theory \eqref{eq:HStripletactionlevel1} for the bulk fields $\varphi_{\mu\nu},C_\mu,D$ and the boundary field $M^{(0)}$ (with the boundary auxiliary fields $N^{(0)},N^{(1)}$ integrated out) which is accompanied by a decoupled massless spin-one theory of the fields $a_\mu$ and $a$. These results for the massive level-1 open-string fields thus provide first evidence that the gauge-invariant actions presented in this paper are not just artifacts appearing in the massless sector and the tensionless limit of SFT, but that they can be extended to cover also the massive levels of the full SFT.


After achieving a gauge-invariant formulation of the full free SFT on a target space with boundary, it will be time to understand how the presence of the boundary is reflected in the structure of the interaction terms. The results which we present in Appendix \ref{app:nilpotent} suggest that it may be possible to achieve this in the (appropriately extended) framework of cyclic homotopy algebras. We expect that a complete understanding of boundary terms in SFT will provide new exciting applications in several directions including \cite{Mazel:2024alu, Scheinpflug:2023lfn, Maccaferri:2023gof, Cho:2023mhw, Kim:2024dnw}.


\section*{Acknowledgments}
We thank Minjae Cho, Harold Erbin, Atakan Firat, Edward Mazenc, Martin Schnabl and Georg Stettinger for useful discussions. JV also thanks CEICO, Institute of Physics, Czech Academy of Sciences for their hospitality during the completion of this work. The work of CM and AR is partially supported by the MIUR PRIN contract 2020KR4KN2 “String Theory as a bridge between Gauge Theories and Quantum Gravity” and by the INFN project ST$\&$FI “String Theory and Fundamental Interactions”. RP acknowledges the financial support from the European Structural and Investment Funds and the Czech Ministry of Education, Youth, and Sports (project FORTE CZ.02.01.01/00/22\_008/0004632). The work of JV is supported by the ERC Starting Grant 853507.

\appendix

\section{Linearizing gravity around flat space with a flat boundary}\label{app:linear}

Let us now explore in detail how the gauge-invariant massless spin-2 actions on a flat half-space with a flat boundary which we presented in Sections \ref{sec:massless} and \ref{sec:tensionless} arise from linearizing the Einstein-Hilbert action of general relativity supplemented by the Gibbons-Hawking-York boundary term. 

\subsection{Rigid boundary}

We will start our presentation by computing the second variation of the pure-gravity action
\begin{align}
    S_\mathrm{GR}= S_\mathrm{EH}+S_\mathrm{GHY}= \int_M d^d x\,\sqrt{g}\, R +\int_{\p M} d^{d-1} y\,\sqrt{\gamma}\, 2K\,.\label{eq:actionGRApp}
\end{align}
with respect to the bulk metric $g_{\mu\nu}$, keeping the boundary $\p M$ of the (topological) manifold $M$ rigid. Upon substituting for the background to be that of a flat half-space with a flat boundary, we will see this reproduces the graviton part of the massless spin-2 action \eqref{eq:HStripletactionlevel1} with the boundary field $M^{(0)}$ gauged away. A fully-gauge invariant action involving $M^{(0)}$ will then be recovered below in Section \ref{subsec:fluct} where variations in the shape of $\p M$ are taken into play.

Let us begin by considering the first variation of \eqref{eq:actionGRApp} with respect to the bulk metric $g_{\mu \nu}$, namely
\begin{equation}
    \delta_g S_\mathrm{GR} = \int_M d^d x\, \delta_g(\sqrt{g}R) + \int_{\partial M} d^{d-1}y\, \delta_g(\sqrt{\gamma}\,2K)\,.
\end{equation}
We will mostly follow the detailed exposition of \cite{bavera}. Performing standard manipulations, one first finds that the variation of the Einstein-Hilbert lagrangian takes the form
\begin{align}
    \delta_g(\sqrt{g} R) &= \sqrt{g}\,\bigg(\frac{1}{2}R g^{\alpha \beta}-R^{\alpha \beta}\bigg)\,\delta g_{\alpha \beta}+ \sqrt{g}\,\nabla_\mu \big(g^{\alpha \beta}  \delta_g \Gamma_{\alpha \beta}^{\mu}-g^{\alpha \mu}\delta_g \Gamma_{\beta \alpha}^{\beta}\big)\,.\label{eq:first_var_0}
\end{align}
meaning that the r.h.s.\ of \eqref{eq:first_var_0} becomes 
\begin{align}
    \delta_g S_\mathrm{GR} &=
      \int_M d^d x \,\sqrt{g}\,\Big(\tfrac{1}{2}R g^{\alpha \beta}-R^{\alpha \beta}\Big)\,\delta g_{\alpha \beta} +\nonumber\\
      &\hspace{1cm}
    + \int_{\partial M} d^{d-1} y \sqrt{\gamma}\, n_\mu \Big(g^{\alpha \beta} \delta_g \Gamma_{\alpha \beta}^\mu - g^{\alpha \mu }\delta_g \Gamma_{\beta \alpha}^\beta\Big) 
     + \int_{\partial M} d^{d-1}y\, \delta_g(\sqrt{\gamma}\,2K)\,.\label{eq:first_var_1}
\end{align}
As expected, the first variation is proportional to the Einstein field equations up to a boundary term.

\subsubsection{Essential hypersurface toolkit}

Our goal is now to show that we can rewrite the second line of \eqref{eq:first_var_1} in terms of the symmetric extrinsic curvature
\begin{align}
    K_{\alpha\beta} = \nabla_\alpha n_\beta - n_\alpha a_\beta\,,
\end{align}
where we have introduced the acceleration 1-form $a_\beta=n^\lambda\nabla_\lambda n_\beta$ which is tangential to $\p M$. The induced metric $\gamma_{ab}$ on the boundary can be introduced as
\begin{align}
    \gamma_{ab} = e^\alpha_a\, e^\beta_b\, g_{\alpha\beta}\,,
\end{align}
where, defining the boundary in terms of an embedding map $x^\alpha(y^a)$, the transformation matrices $e^\alpha_a$ can be defined as
\begin{align}
    e^\alpha_a = \frac{\p x^\alpha(y)}{\p y^a}\,.
\end{align}
At the same time, these give the components of the $d-1$ vector fields which are tangent to $\p M$. The inverse induced metric $\gamma^{ab}$ also satisfies
\begin{align}
    e^\alpha_a\, e^\beta_b \gamma^{ab} = g^{\alpha\beta}-n^\alpha n^\beta \coloneqq \gamma^{\alpha\beta}\,.
\end{align}
The trace $K$ of $K_{\alpha\beta}$ can then be expressed as 
\begin{align}
    K = g^{\alpha\beta}K_{\alpha\beta} =\gamma^{\alpha\beta}K_{\alpha\beta}=\nabla_\alpha n^\alpha\,,
\end{align}
since $n^{\alpha}K_{\alpha\beta}=n^{\alpha}a_\alpha=0$. Furthermore, note that defining the hypersurface $\p M$ implicitly as
\begin{align}
    F(x)=0\label{eq:implicit_F}
\end{align}
for some function $F$ on $M$, the expression 
\begin{align}
    n_\alpha=\frac{\p_\alpha F}{\sqrt{g^{\rho\sigma}\p_\rho F \p_\sigma F}}\,.\label{eq:nF}
\end{align}
defines the unit form field normal to $\p M$. Finally, the \emph{intrinsic} covariant derivative of a one-form field $A_a=e^\alpha_a A_\alpha$ where $A_\alpha$ is tangential to $\p M$ (that is $n^\alpha A_\alpha=0$) can be defined as
\begin{align}
    D_a A_b  = e^\alpha_a e^\beta_b \nabla_\alpha A_\beta\,.
\end{align}
This can be shown to coincide with the covariant derivative on $\p M$ defined in terms of the Levi-Civita connection compatible with the induced metric $\gamma_{ab}$ which is used to raise and lower latin indices. Given $D_a A_b$, let us also define the induced covariant derivative
\begin{align}
    D^\alpha A^\beta \coloneqq e^\alpha_a e^\beta_b D^a A^b  = \tensor{\gamma}{^\alpha_\mu}\, \tensor{\gamma}{^\beta_\nu} \nabla^\mu A^\nu\,.\label{eq:Dab}
\end{align}
In particular, we note the relation
\begin{align}
    D_\alpha A^\alpha = g_{\alpha\beta} e^\alpha_a e^\beta_b D^a A^b = \gamma_{ab} D^a A^b = D_a A^a
\end{align}
between the induced and the intrinsic covariant divergences.

\subsubsection{Boundary terms of the first variation}
 
Given this preparation, we first notice that we can rewrite the variations of the Levi-Civita connection appearing in \eqref{eq:first_var_1} as
\begin{equation}
        n_\mu\left(g^{\alpha \beta} \delta_g \Gamma_{\alpha \beta}^\mu - g^{\alpha \mu }\delta_g \Gamma_{\beta \alpha}^\beta\right) = \nabla_\alpha \,\delta u^\alpha - 2\,\delta_g K + \nabla_\alpha n_\beta \,\delta_g g^{\alpha \beta} \label{eq:first_var_3}
\end{equation} 
where we have introduced a tangent vector field
\begin{equation}
    \delta u^\alpha = \delta_g n^\alpha + g^{\alpha \beta} \delta_g n_\beta\,,
\end{equation}
and recalled the form $K=\nabla_\mu n^\mu$ for the trace of the extrinsic curvature tensor of $\p M$.
At the same time, after some straightforward manipulations,
we can rewrite
\begin{equation}
D_a \delta u^a =  D_\alpha \delta u^\alpha 
         = \nabla_\alpha \delta u^\alpha + \delta u^\alpha a_\alpha\,.
\end{equation}
This enables us to further rewrite the r.h.s.\ of \eqref{eq:first_var_3} as
\begin{equation}
    n_\mu\Big(g^{\alpha \beta} \delta_g \Gamma_{\alpha \beta}^{\mu}-g^{\alpha \mu}\delta_g \Gamma_{\beta \alpha}^{\beta}\Big) = D_a \delta u^a - 2\,\delta_g K - \delta u^\alpha\, a_\alpha  + \nabla_\alpha n_\beta\,\delta_g g^{\alpha\beta}\,.
\end{equation}
Finally,
we can compute
\begin{equation}
\begin{split}
    \delta u^\alpha \,a_\alpha 
    & = a_\alpha n_\beta \,\delta_g g^{\alpha \beta} + 2a^\beta \delta n_\beta\,,\label{eq:ua}
\end{split}
\end{equation}
where, assuming that the boundary is rigid, the variation $\delta_g n_\beta$
can be expressed as (recalling the form \eqref{eq:nF} of $n_\beta$)
\begin{equation} \label{eq: deltan}
    \delta_g n_\beta 
     = -\frac{1}{2} n_\beta \,\frac{\partial_\rho F\, \partial_\sigma F \,\delta g^{\rho \sigma}}{\partial_\rho F \,\partial_\sigma F g^{\rho \sigma}} \,.
\end{equation}
As we can see that $\delta_g n_\beta$ is parallel to $n_\beta$, we conclude that the last term of \eqref{eq:ua} vanishes.
Hence, we can finally rewrite
\begin{equation}
        n_\mu\Big(g^{\alpha \beta} \delta_g \Gamma_{\alpha \beta}^{\mu}-g^{\alpha \mu}\delta_g \Gamma_{\beta \alpha}^{\beta}\Big) 
       = D_a \delta u^a - 2\,\delta_g K  + K_{\alpha \beta}\,\delta_g g^{\alpha \beta}\label{eq:first_var_4}
\end{equation}
where we recognized the extrinsic curvature tensor $K_{\alpha \beta} = \nabla_\alpha n_\beta - n_\alpha a_\beta$. Substituting the result \eqref{eq:first_var_4} back into \eqref{eq:first_var_1}, we therefore find
\begin{align}
          \delta_g S_\mathrm{GR} &=
    \int_M d^d x \,\sqrt{g}\,\Big(\tfrac{1}{2}R g^{\alpha \beta}-R^{\alpha \beta}\Big)\,\delta g_{\alpha \beta} +\nonumber\\
    &\hspace{0.4cm}
     - \int_{\partial M} d^{d-1} y \sqrt{\gamma} \, K^{\alpha \beta}\,\delta g_{\alpha\beta} 
     + \int_{\partial M} d^{d-1}y\, \delta_g(\sqrt{\gamma})\, 2K + \int_{\partial M} d^{d-1} y \,\sqrt{\gamma}\, D_a \delta u^a \,,
\end{align}
where we will assume that the last term can be dropped without picking any contributions from the boundaries of $\p M$ (corners of $M$). It remains to express
\begin{align}
    \delta_g\sqrt{\gamma} = \frac{1}{2}\sqrt{\gamma}\,\gamma^{ab}\delta_g\gamma_{ab}=\frac{1}{2} \sqrt{\gamma}\,\gamma^{ab} e^\alpha_a e^\beta_b
    \delta g_{\alpha\beta}=\frac{1}{2} \sqrt{\gamma}\,\gamma^{\alpha\beta} 
    \delta g_{\alpha\beta}\,,
\end{align}
where, since the boundary $\p M$ is being kept rigid, we have used that $\delta_g e_a^\alpha=0$. 
Therefore, the first variation of the action \eqref{eq:actionGRApp} with respect to the bulk metric $g_{\mu\nu}$ takes the final form
\begin{align}
    \delta_g S_\mathrm{GR} = \int_M d^d x\,\sqrt{g}\,\big(\tfrac{1}{2}Rg^{\alpha\beta}-R^{\alpha\beta}\big)\,\delta g_{\alpha\beta}+\int_{\p M} d^{d-1}y\,\sqrt{\gamma}\,(K\gamma^{\alpha\beta}-K^{\alpha\beta})\,\delta g_{\alpha\beta}\,.\label{eq:first_var}
\end{align}

\subsubsection{Second variation in flat half-space}

We can now proceed with varying the action with respect to the bulk metric once more. In order not to get overwhelmed with complicated expressions and, at the same time, having in mind our final goal of finding an action for massless spin-2 fluctuations around a flat half-space bounded by a flat boundary, we will be substituting all bulk and boundary quantities with their flat-space values as we go. In particular, using the cartesian coordinates $x^\mu=(y^a,z)$ introduced in Section \ref{subsec:init},
we will be setting
\begin{subequations}
\label{eq:flat_background}
    \begin{align}
         \overline{g}_{\mu \nu} &= \eta_{\mu \nu} = \text{diag}[+1,\cdots,+1]\,,\\
         \overline{R} &= 0\,,\\
         \overline{K} &= 0\,,\\
         \overline{n}_\mu &=(0,\ldots,0,1)=\delta_\mu^z\,.
\end{align}
\end{subequations}
The flat half-space $M$ will be taken to extend for $z<0$. Notice that this is indeed a consistent critical background as it makes the first variation \eqref{eq:first_var} vanishing.

Let us start by writing down the second variation as
\begin{align}
        \delta^2_g S_\mathrm{GR}\big|_\text{flat} 
        &= \int_M d^d x \,\Big(\tfrac{1}{2}\eta^{\alpha \beta}\delta_g R - \delta_g R^{\alpha \beta}\Big)\, \delta g_{\alpha \beta} + \nonumber\\
        &\hspace{4cm}+\int_{\partial M} d^{d-1}y \,\Big[ (\eta^{\alpha \beta}-\delta^\alpha_z\delta^\beta_z)\delta_g K - \delta_g K^{\alpha \beta} \Big] \,\delta g_{\alpha \beta}\,,\label{eq:second_var_0}
\end{align}
where we have noted that the variation has to hit the curvatures, otherwise we get a zero upon substituting for the background values \eqref{eq:flat_background}. To evaluate the r.h.s.\ of \eqref{eq:second_var_0}, we have to compute the following
\begin{subequations}
\label{eq:R_var}
\begin{align}
     \delta_g R \big|_\text{flat} &= \partial^\rho \partial^\sigma \delta g_{\rho \sigma} - \eta^{\rho \sigma} \partial^2  \delta g_{\rho \sigma}\,, \\[1mm]
     \delta_g R^{\alpha \beta} \big|_\text{flat} &= \frac{1}{2}\eta^{\alpha \nu }\eta^{\beta \sigma} \Big(  \partial^\rho \partial_\nu \delta g_{\rho \sigma} + \partial^\rho \partial_\sigma \delta g_{\rho \nu} - \partial^2 \delta g_{\nu \sigma } - \partial_\sigma \partial_\nu \eta^{\lambda \mu} \delta g_{\lambda \mu}  \Big)\,, 
\end{align}
\end{subequations}
as well as
\begin{align}
    \delta_g n_\beta \big|_\text{flat} &= -\frac{1}{2}\delta_\beta^z \,\delta g^{zz}  = +\frac{1}{2} \delta_\beta^z \,\delta g_{zz}\,,
\end{align}
where we have used \ref{eq: deltan} to derive the last result.
Notice that $\delta_g n_\beta|_\text{flat}$ is non-vanishing only along the $z$ component.
Besides that, we will need the result
\begin{equation}
       \delta_g a_\beta \big|_\text{flat}  
        = \delta^\nu_z\big( \partial_\nu \delta_g n_\beta- \delta_\rho^z\,\delta_g \tensor{\Gamma}{^\rho_{\nu \beta}} \big)
        = \frac{1}{2} \delta_\beta^z \partial_z \delta g_{zz} - \delta_g \Gamma_{z\beta}^z 
        = \frac{1}{2} \delta_\beta^z \partial_z \delta g_{zz} - \frac{1}{2} \partial_\beta \delta g_{zz}\,.
\end{equation}
This enables us to compute the variation of the extrinsic curvature, namely
\begin{subequations}
\begin{align}
        \delta_g K_{\alpha \beta} \big|_\text{flat} 
        & = \partial_\alpha \delta_g n_\beta - \delta_g \Gamma_{\alpha \beta}^\rho \delta_\rho^z - \delta_\alpha^z \delta_g a_\beta \\
        & = \frac{1}{2} \delta_\beta^z \partial_\alpha \delta g_{zz} - \delta_g \Gamma_{\alpha \beta}^z - \frac{1}{2} \delta_\alpha^z \big( \delta_\beta^z \partial_z \delta g_{zz} - \partial_\beta \delta g_{zz} \big)\,,
    \end{align}
\end{subequations}
or, equivalently,
\begin{subequations}
\label{eq:Kab_var}
\begin{align}
    & \delta_g K_{ab} \big|_\text{flat} = -\frac{1}{2}\big(\partial_a \delta g_{zb} + \partial_b \delta g_{za}- \partial_z \delta g_{ab}\big)\,,\\
    & \delta_g K_{zz} \big|_\text{flat} = 0\,, \\[1mm]
    & \delta_g K_{za} \big|_\text{flat} = 0\,,
\end{align}
\end{subequations}
as well as the variation of its trace
\begin{equation}
    \delta_g K \big|_\text{flat} = -\partial^b \delta g_{zb} + \frac{1}{2} \eta^{ab} \partial_z \delta g_{ab}\,.\label{eq:K_var}
\end{equation}
Putting together the results \eqref{eq:Kab_var} and \eqref{eq:K_var}, we can, after some algebra, express the integrand of the boundary term in \eqref{eq:second_var_0} as 
\begin{align}
    &\big[ (\eta^{\alpha \beta}-\delta^\alpha_z \delta^\beta_z)\delta_g K  -  \delta_g K^{\alpha \beta}\big] \,\delta g_{\alpha \beta} =\big( \eta^{ac} \eta^{bd}-\eta^{ab} \eta^{cd}\big)\Big(\partial_c \delta g_{za} - \frac{1}{2}\partial_z \delta g_{cd}\Big)\,\delta g_{ab}\,.\label{eq:bndy_eom_var}
\end{align}
At this point, let us identify the variation $\delta g_{\alpha\beta}$ with the linearized fluctuation $h_{\alpha\beta}$ of the bulk metric, namely
\begin{align}
    h_{\alpha\beta} \coloneqq \delta g_{\alpha\beta}\,.\label{eq:hdg}
\end{align}
Substituting then for the bulk curvature variations \eqref{eq:R_var}, as well as for the boundary-term variation \eqref{eq:bndy_eom_var} into \eqref{eq:second_var_0}, we have
\begin{align}
    \frac{1}{2}\delta^2_g S_\mathrm{GR}\big|_\text{flat}(h_{\alpha\beta}) 
        &= \frac{1}{4}\int_M d^d x \,h_{\alpha \beta}\,\Big(\eta^{\alpha \beta}(\partial^\rho \partial^\sigma \delta g_{\rho \sigma} - \eta^{\rho \sigma} \partial^2  \tensor{h}{^\lambda_\lambda})   -2\partial_\lambda \partial^\alpha h^{\lambda \beta} +\nonumber\\
        &\hspace{7.3cm}+ \partial^2 h^{\alpha\beta } + \partial^\alpha \partial^\beta  \tensor{h}{^\lambda_{\lambda} }  \Big) + \nonumber\\
        &\hspace{2cm}+\frac{1}{4}\int_{\partial M} d^{d-1}y\,h_{ab}\,\big( \eta^{ac} \eta^{bd}-\eta^{ab} \eta^{cd}\big)\Big(2\partial_c h_{za} - \partial_z h_{cd}\Big)\,.\label{eq:second_var_2}
\end{align}
We notice that 1.\ in the bulk term of \eqref{eq:second_var_2}, there are a number of terms containing second derivatives with respect to the $z$ coordinate, and, 2.\ in the boundary term of \eqref{eq:second_var_2}, there are terms with first derivatives with respect to $z$. These would render the variational principle given by the action \eqref{eq:second_var_2} inconsistent. Luckily though, all these turn out to cancel among themselves upon integrating some of the derivatives in the bulk terms by parts. Indeed, after a few straightforward steps, the action \eqref{eq:second_var_2} becomes
\begin{align}
    &\frac{1}{2}\delta_g^2 S_\mathrm{GR}\big|_\text{flat}(h_{\alpha\beta}) =\nonumber\\
    &\hspace{0.4cm}=-\frac{1}{2}\int_M d^d x\,\bigg(\frac{1}{2}\p^\lambda h^{\alpha\beta}\p_\lambda h_{\alpha\beta}-\frac{1}{2}\p^\lambda \tensor{h}{^\rho_\rho}\p_\lambda \tensor{h}{^\sigma_\sigma}-\p^{\alpha}h^{\lambda\beta}\p_\lambda h_{\alpha\beta}+\p^\beta \tensor{h}{^\rho_\rho}\p^\alpha h_{\alpha\beta}\bigg)+\nonumber\\
    &\hspace{7.9cm}-\frac{1}{2}\int_{\p M} d^{d-1}y\, h_{zd} \,\p^d(2h_{zz}-\tensor{h}{^\rho_\rho})\,,\label{eq:second_var}
\end{align}
which constitutes a well-defined variational principle for linearized massless spin-2 fluctuations $h_{\alpha\beta}$ on the given background. Furthermore, \eqref{eq:second_var} is invariant under the linearized gauge transformation
\begin{align}
   \delta_\xi h_{\alpha\beta} = \p_\alpha \xi_\beta +\p_\beta\xi_\alpha\label{eq:gauge_app}
\end{align}
provided that we restrict the gauge parameter $\xi_\alpha$ to satisfy
\begin{align}
    n^\alpha \xi_\alpha = \xi_z =0\label{eq:gauge_rest}
\end{align}
at $\p M$. As the gauge transformation \eqref{eq:gauge_app} descends from the linearized trasformation of the bulk metric under the infinitesimal diffeomorphism
\begin{align}
    x^{\mu} \longrightarrow (x')^{\mu} = x^\mu-\xi^\mu\,,\label{eq:diffeo}
\end{align}
the meaning of \eqref{eq:gauge_rest} is clear: we are confining ourselves to considering diffeomorphisms which preserve the hyperplane at $z=0$. The action \eqref{eq:second_var} matches with the massless spin-2 actions obtained at level 0 of the closed string and level 1 of the tensionless open string upon setting both the boundary degree of freedom $l(y^a)$ and the dilaton $\phi(x^\mu)$ to zero.

\subsection{Fluctuating boundary}
\label{subsec:fluct}

We will now show that the \emph{whole} gauge-invariant massless spin-2 action (including non-zero boundary field $l(y^a)$) obtained from SFT can be recovered by linearizing GR around a flat half-space: we will see that the field $l(y^a)$ should be interpreted as being proportional to a transverse displacement of the boundary.

\subsubsection{Parametrization of $\p M$}

The boundary $\p M$ can be generally described by an implicit equation of the form \eqref{eq:implicit_F} in terms of a function $F$ on $M$. Keeping in mind the concrete example of linearized fluctuations around a flat hyperplane bounding a half-space inside $\mathbb{R}^d$, we will consider $F$ to have the particular form
\begin{align}
    F(x^\mu) = z-f(y^a)\,.\label{eq:Ff}
\end{align}
Here, as in the main body of the paper, $y^a$ are some coordinates along the hyperplane while $z$ is the coordinate running in the normal direction. The function $f(y^a)$ then parametrizes a class of smooth deformations of the hyperplane at $z=0$. 
Upon enacting the infinitesimal diffeomorphism \eqref{eq:diffeo} of $M$, the function $f(y^a)$ obeys the linearized transformation law (dropping $\mathcal{O}(\xi f)$ terms)
\begin{align}
    \delta_\xi f(y^a) = -\xi_z(y^a)\,,\label{eq:f_diffeo}
\end{align}
where we have used the shortcut $\xi_z(y^a)=\xi_z(x^\mu(y^a))$. 
The hypersurface $\p M$ given by \eqref{eq:Ff} can be alternatively described by the embedding map
\begin{align}
    x^\mu(y^c) = \delta_b^\mu \,y^b+\delta^\mu_z f(y^c)
    \,.
\end{align}
When we vary with respect to $f$, the embedding map then varies as
\begin{align}
    \delta_f \big(x^\mu(y^c)\big ) = \delta^\mu_z\, \delta f(y^c)\,.\label{eq:var_emb}
\end{align}
Correspondingly, for the transition maps $e^\mu_a$ and their variations $\delta_f e^\mu_a$, we can write
\begin{subequations}
\begin{align}
       e^\mu_a(y^c) &=\delta_a^\mu +\delta^\mu_z \,\p_a f(y^c)\,,\\
       \delta_f\big(e^\mu_a(y^c)\big)&=\delta^\mu_z \,\p_a \delta f(y^c)\,.\label{eq:var_e}
\end{align}
\end{subequations}

\subsubsection{Linearized action: summary}

Given this setup, the action \eqref{eq:actionGRApp} can be considered as a functional of both the bulk metric $g_{\mu\nu}$ as well as of the function $f(y^a)$ parametrizing the boundary manifold $\p M$. Varying the action with respect to $f(y^a)$ will therefore yield additional terms into the linearized action \eqref{eq:second_var} for massless spin-2 fluctuations. These will involve the displacement $\delta f(y^a)$ of the boundary. Introducing a rescaled fluctuation
\begin{align}
  l(y^a)\coloneqq -2\,  \delta f(y^a) \label{eq:fl}
\end{align}
and given the transformation property \eqref{eq:f_diffeo} of $f(y^a)$, we can observe that the field $l(y^a)$ transforms as
\begin{align}
    \delta_\xi l(y^a) = 2\,\xi_z (y^a)\,.
\end{align}
This exactly matches the gauge transformation of the corresponding boundary degree of freedom entering the massless spin-2 actions derived from SFT. Indeed, we will see below that varying the GR action \eqref{eq:actionGRApp} once with respect to $f$, once with respect to $g_{\alpha\beta}$, substituting for the flat background $g_{\mu\nu} = \eta_{\mu\nu}$ with flat boundary $f(y^a)=0$ and, finally, using the redefinitions \eqref{eq:hdg}, \eqref{eq:fl}, we will obtain a boundary term which 1.\ restores the full invariance of the action \eqref{eq:second_var} under the gauge transformation \eqref{eq:gauge_app} and 2.\ exactly agrees with the term proportional to $l(y^a)$ in the linearised SFT actions \eqref{eq:level0closedactionfinal2} and \eqref{eq:HStripletactionlevel1} upon setting the dilaton $\phi=0$. That is to say, we will show that
\begin{subequations}
\begin{align}
   \frac{1}{2} \delta_f( \delta_g S_\mathrm{GR})\big|_\text{flat}(h_{\alpha\beta},l)&=\frac{1}{2}\delta_g( \delta_f S_\mathrm{GR}) \big|_\text{flat}(h_{\alpha\beta},l) \\[2mm]
    &= -\frac{1}{4}\int_{\p M} d^{d-1}y\,\big(h^{ab}-\eta^{ab}\tensor{h}{^c_c}\big)\,\p_a\p_b l\,.\label{eq:var_fg_S}
\end{align}
\end{subequations}
The interchangeability of the variations with respect to $g_{\mu\nu}$ and $f$ was of course to be expected since the action \eqref{eq:actionGRApp} is a smooth functional of $g_{\mu\nu}$ and $f$.
We will also show that 
\begin{align}
     \delta_f^2 S_\mathrm{GR}\big|_\text{flat}(h_{\alpha\beta},l)=0\,,
\end{align}
meaning that all terms in the linearized action which are quadratic in the field $l(y^a)$ vanish. Altogether, the full linearized action therefore reads
\begin{subequations}
\label{eq:lin_GR_final}
\begin{align}
    &S(h_{\alpha\beta},l)=\nonumber\\[2mm]
    &\hspace{0.2cm}=\frac{1}{2}\delta_g^2 S_\mathrm{GR}\big|_\text{flat}(h_{\alpha\beta})+ \delta_f \delta_g S_\mathrm{GR}\big|_\text{flat}(h_{\alpha\beta},l)+\frac{1}{2} \delta_f^2 S_\mathrm{GR}\big|_\text{flat}(h_{\alpha\beta},l)\\[1mm]
    &\hspace{0.2cm}=-\frac{1}{2}\int_M d^d x\,\bigg(\frac{1}{2}\p^\lambda h^{\alpha\beta}\p_\lambda h_{\alpha\beta}-\frac{1}{2}\p^\lambda \tensor{h}{^\rho_\rho}\p_\lambda \tensor{h}{^\sigma_\sigma}-\p^{\alpha}h^{\lambda\beta}\p_\lambda h_{\alpha\beta}+\p^\beta \tensor{h}{^\rho_\rho}\p^\alpha h_{\alpha\beta}\bigg)+\nonumber\\[-1mm]
    &\hspace{3.7cm}-\frac{1}{2}\int_{\p M} d^{d-1}y\,\Big[ h_{zd} \,\p^d(2h_{zz}-\tensor{h}{^\rho_\rho})-\big(\eta^{ab}\tensor{h}{^c_c}-h^{ab}\big)\,\p_a\p_b l\Big]\,.
\end{align}
\end{subequations}
This is in exact agreement with the pure-graviton part of both the closed SFT action \eqref{eq:level0closedactionfinal2}, as well as the tensionless open SFT action \eqref{eq:HStripletactionlevel1}.

\subsubsection{Linearized action: detailed calculation}

Let us start by varying the action \eqref{eq:actionGRApp} once with respect to $f$. Recall that since $f$ was defined to be a function of $y$ only, the variation $\delta f$ does not depend on the variable $z$.

Collecting first the necessary ingredients, we start by computing the variation of the induced metric 
\begin{align}
    \delta_f\gamma_{ab}(y) = \delta_f\big(e_a^\alpha(y)\, e_b^\beta(y) \,g_{\alpha\beta}(x(y))\big)\,.
\end{align}
Substituting from \eqref{eq:var_emb} and \eqref{eq:var_e} for the variations of the embedding functions and the transition matrices $e_a^\alpha$, we end up with
\begin{align}
    \delta_f\gamma_{ab} = \delta f \, e_a^\alpha\, e_b^\beta\,\p_z g_{\alpha\beta}+ \big(e_a^\alpha \p_b \delta f+e_b^\alpha \p_a \delta f\big) \,g_{\alpha z}\,.
\end{align}
It follows that for the variation of the determinant of the induced metric, we can write
\begin{align}
    \delta_f \sqrt{\gamma} = \frac{1}{2}\sqrt{\gamma}\Big[\delta f \, \gamma^{\alpha\beta} \p_z g_{\alpha\beta}+ \gamma^{ab}\big(e_a^\alpha \p_b \delta f+e_b^\alpha \p_a \delta f\big) \,g_{\alpha z}\Big]\,.
\end{align}
Notice that when we substitute for the flat background \eqref{eq:flat_background} (which, in particular, amounts to setting $f=0$), both $\delta_f\gamma_{ab}$ and $\delta_f \sqrt{\gamma}$ vanish, that is
\begin{subequations}
    \begin{align}
        \delta_f\gamma_{ab}\big|_\text{flat} = 0\,,\label{eq:var_f_gamma_flat}\\
        \delta_f \sqrt{\gamma} \big|_\text{flat} = 0\,.
    \end{align}
\end{subequations}
We will also need to vary the unit form $n_\mu$ normal to $\p M$. Varying the r.h.s.\ of \eqref{eq:nF}, we obtain
\begin{align}
    \delta_f n_\mu (x(y)) = -\frac{\delta_\mu^a\p_a \delta f}{\sqrt{g^{\rho\sigma}\p_\rho F \p_\sigma F}}+ n_\mu\bigg(\frac{g^{\rho a}\p_\rho F\, \p_a \delta f}{g^{\rho\sigma}\p_\rho F \p_\sigma F}-\frac{1}{2}\delta f\, \frac{\p_z g^{\rho\sigma}\p_\rho F \p_\sigma F}{g^{\rho\sigma}\p_\rho F \p_\sigma F}\bigg)\,.
\end{align}
In the flat background \eqref{eq:flat_background}, this gives simply
\begin{align}
    \delta_f n_\mu (x(y))\big|_\text{flat} = -\p_\mu \delta f\,.
\end{align}
Finally, the trace of the extrinsic curvature varies with $f$ as
\begin{align}
    \delta_f K(x(y)) &= \nabla_\mu\bigg(n^\mu \frac{g^{\rho a}\p_\rho F\, \p_a \delta f}{g^{\rho\sigma}\p_\rho F \p_\sigma F} -\frac{\delta_a^\mu\p^a \delta f}{\sqrt{g^{\rho\sigma}\p_\rho F \p_\sigma F}} \bigg)+\nonumber\\
    &\hspace{1cm}+\delta f\,\bigg(\p_z g^{\mu\nu}\,\nabla_\mu n_\nu- g^{\mu\nu}\p_z\Gamma^\rho_{\mu\nu}\,n_\rho \bigg) -\frac{1}{2}\nabla_\mu\bigg(n^\mu\delta f\, \frac{\p_z g^{\rho\sigma}\p_\rho F \p_\sigma F}{g^{\rho\sigma}\p_\rho F \p_\sigma F}\bigg)\,.\label{eq:var_f_K}
\end{align}
This reduces to
\begin{align}
    \delta_f K(x(y))\big|_\text{flat} &= -\p^a \p_a \delta f
\end{align}
in the flat half-space.

Hence, varying the action \eqref{eq:actionGRApp} once with respect to $f$, we obtain
\begin{align}
    \delta_f S_\mathrm{GR} &= \int_{\p M} d^{d-1}y\,\sqrt{g}\, R\,\delta f+\int_{\p M} d^{d-1}y\,\sqrt{\gamma}\, \big(K\,\gamma^{ab} \delta_f\gamma_{ab}+2\delta_f K\big)\label{eq:first_f_var}
\end{align}
where in the first term, we have remembered that the limits of the bulk (Einstein-Hilbert) integral vary with $f$. We have also concluded that explicitly substituting for the variations $\delta_f\gamma_{ab}$ and $\delta_f K$ would not have been particularly illuminating at this point. It is nevertheless reassuring to note that evaluating \eqref{eq:first_f_var} in the flat half-space, we obtain
\begin{align}
    \delta_f S_\mathrm{GR}\big|_\text{flat} &= -2\int_{\p M} d^{d-1}y\, \p^a \p_a \delta f =0
\end{align}
assuming, as usual, that the boundary of $\p M$ can be ignored. This means that the flat half-space extremizes the action \eqref{eq:actionGRApp} even with respect to the transverse displacement $f$ of the boundary $\p M$.

Let us now vary with respect to $f$ once more, this time immediately evaluating the results for the flat half-space. First, we can quickly note that
\begin{align}
    \delta_f R(x(y))\big|_\text{flat} = \delta f\,\p_z  R\big|_\text{flat} =0\,,
\end{align}
meaning that the first term in \eqref{eq:first_f_var} will not contribute. For the second term, the only new necessary object to compute is $\delta_f^2 K|_\text{flat}$ because the would-be contribution of $\delta_f^2 \gamma_{ab}|_\text{flat}$ is pre-multiplied by $K$ which kills it. After an uneventful tour through the terms appearing on the r.h.s.\ of \eqref{eq:var_f_K}, we obtain
\begin{align}
    \delta_f^2 K|_\text{flat} = \p_\mu\big(-\delta_z^\mu \p^a \delta f \p_a\delta f \big)=0\,,
\end{align}
where we have recalled that the variation $\delta f$ does not depend on $z$. Hence, since we can write
\begin{align}
    \delta_f^2 S_\mathrm{GR}\big|_\text{flat} &=2\int_{\p M} d^{d-1}y\,\sqrt{\gamma}\, \delta_f^2 K|_\text{flat} =0\,,
\end{align}
we confirm our claim that the linearized free action \eqref{eq:lin_GR_final} does not contain any terms which would be quadratic in the boundary displacement.

Let us proceed with computing the mixed variations evaluated in flat half-space, starting with the $f$-variation of \eqref{eq:first_var}. The only new ingredients needed here are 
\begin{align}
    \delta_f a_\mu(x(y))\big|_\text{flat} = \p_z\big(\delta_f n_\mu|_\text{flat}\big)=-\p_\mu\p_z \delta f =0\,,
\end{align}
which can be used to compute
\begin{align}
    \delta_f K_{\mu\nu}(x(y))\big|_\text{flat} = -\delta_\mu^a \,\delta_\nu^b\, \p_a\p_b \delta f\,.
\end{align}
Furthermore, when we vary the limits of the bulk term in \eqref{eq:first_var}, we get a zero upon substituting for the flat half-space background since it solves the Einstein field equation in the bulk. Hence, we can write
\begin{subequations}
    \begin{align}
       \frac{1}{2} \delta_f(\delta_g S_\text{GR})\big|_\text{flat} &= \frac{1}{2}\int_{\p M} d^{d-1}y\,\big((\eta^{\alpha\beta}-\delta^\alpha_z\delta^\beta_z)\,\delta_f K\big|_\text{flat} - \delta_f K^{\alpha\beta}\big|_\text{flat}\big)\,\delta g_{\alpha\beta}\\
        &= \frac{1}{2}\int_{\p M} d^{d-1}y\,\big(\delta g_{ab}-\eta_{ab}\,\eta^{cd}\delta g_{cd}\big)\p^a \p^b \delta f\,,
    \end{align}
\end{subequations}
so that upon substituting for $\delta g_{ab}$ and $\delta f$ in terms of $h_{ab}$ and $l$, we exactly recover the r.h.s.\ of \eqref{eq:var_fg_S}. It remains to verify that the $g_{\mu\nu}$-variation of \eqref{eq:first_f_var} gives the same result. Since the variations with respect to $f$ and $g_{\mu\nu}$ should commute, this is really just a sanity check. First, we write
\begin{align}
    \frac{1}{2} \delta_g(\delta_f S_\text{GR})\big|_\text{flat} &=\frac{1}{2}\int_{\p M} d^{d-1}y\, \delta_g R\big|_\text{flat} \,\delta f+\int_{\p M } d^{d-1}y\,\delta_g(\sqrt{\gamma}\delta_f K)\big|_\text{flat}\,,\label{eq:var_gf_S}
\end{align}
where we have tacitly used the result \eqref{eq:var_f_gamma_flat}. Since $\delta_g R\big|_\text{flat}$ was already computed in \eqref{eq:R_var}, we only have to find
\begin{subequations}
\begin{align}
    &\delta_g(\sqrt{\gamma}\delta_f K)\big|_\text{flat} =\nonumber\\[2mm]
    &\hspace{0.2cm}=\eta^{\mu\nu}\delta_g\Gamma_{\mu\nu}^a \,\p_a \delta f+\delta g_{\mu a}\,\p^\mu \p^a \delta f - \p_z \delta g_{za} \,\p^a\delta f+\nonumber\\
    &\hspace{1cm}- \frac{1}{2}\p^a \big(\p_a\delta f\, \delta g_{zz}\big) - \delta f\, \eta^{\mu\nu}\,\p_z\delta_g \Gamma_{\mu\nu}^z + \frac{1}{2}\delta f\,\p^2_z\,\delta g_{zz}-\frac{1}{2}\p^a\p_a\delta f\,\eta^{ab}\,\delta g_{ab}\\
    &\hspace{0.2cm}= \Big(\p_z \delta g_{cz}-\tfrac{1}{2}\p_c \delta g_{zz}\Big)\,\p^c \delta f +\frac{1}{2}\delta f\, \eta^{ab}\p_z^2 \delta g_{ab}+\nonumber\\
    &\hspace{3cm}+\p^a\Big(\delta g_{ab}\,\p^b \delta f - \tfrac{1}{2}\p_a \delta f \,\delta g_{zz}-\p_z\delta g_{za}\,\delta f-\tfrac{1}{2}\p_a \delta f\,\eta^{cd}\delta g_{cd} \Big)\,,
\end{align}
\end{subequations}
where the very last term, since it is a total derivative along $\p M$, will drop upon substituting into \eqref{eq:var_gf_S}. In total, we obtain
\begin{align}
    &\frac{1}{2} \delta_g(\delta_f S_\text{GR})\big|_\text{flat} =\nonumber\\
    &\hspace{0.1cm}=\frac{1}{2}\int_{\p M} d^{d-1}y\, \big(
    \p^a\p^b \delta g_{ab}+2\p^a\p^z \delta g_{az}-\eta^{ab}\p^c\p_c\delta g_{ab}-\eta^{ab}\p_z^2 \delta g_{ab}-\p^a\p_a \delta g_{zz}
    \big) \,\delta f+\nonumber\\
   &\hspace{4.2cm} +\frac{1}{2}\int_{\p M } d^{d-1}y\,\bigg[  \Big(2\p_z \delta g_{cz}-\p_c \delta g_{zz}\Big)\,\p^c \delta f + \eta^{ab}\p_z^2 \delta g_{ab}\, \delta f\bigg]\,,
\end{align}
where we can observe exact cancellation among all terms containing derivatives with respect to $z$. After straightforward manipulations, we end up with 
\begin{align}
   \frac{1}{2} \delta_g(\delta_f S_\text{GR})\big|_\text{flat}= \frac{1}{2}\int_{\p M} d^{d-1}y\,\big(\delta g_{ab}-\eta_{ab}\,\eta^{cd}\delta g_{cd}\big)\p^a \p^b \delta f=\frac{1}{2} \delta_f(\delta_g S_\text{GR})\big|_\text{flat}\,,
\end{align}
thus confirming our claim \eqref{eq:var_fg_S} and hence, finally, the result \eqref{eq:lin_GR_final}.

\section{Cyclic nilpotent structure for a flat boundary}\label{app:nilpotent}

In this appendix we will aim to demonstrate that the gauge invariance of the action \eqref{eq:TheAction} is a consequence of the fact that \eqref{eq:TheAction} can be rewritten in terms of a nilpotent kinetic operator and a symplectic form with respect to which this operator is cyclic.

Before starting, let us first re-package both the dynamical bulk string field $\Psi(\hat{x})$ and the boundary dynamical string fields $\Sigma_0(\hat{y})$, $\Sigma_1(\hat{y})$ as
\begin{subequations}
\label{eq:sf_repackaging}
\begin{align}
    \Psi^{[1]}(\hat{x}) &= \Sigma_0(\hat{y})+\hat{z}\,\Sigma_1(\hat{y})+\frac{\hat{z}^2}{2!}\Sigma_2(\hat{y})+\ldots =\sum_{k=0}^\infty \frac{\hat{z}^k}{k!}\Sigma_k(\hat{y})\,,\label{eq:zexp}\\
    \Psi^{[2]}(\hat{x}) &= \Psi(\hat{x})\,.
\end{align}
\end{subequations}
Notice that in the definition of $\Psi^{[1]}(\hat{x})$, we had to accompany $\Sigma_0$ and $\Sigma_1$ with additional string fields $\Sigma_2,\Sigma_3,\ldots$ These can be chosen arbitrarily as they will not end up contributing into the action and therefore will not participate in the dynamics. Their sole purpose is to populate the remaining coefficients in the Taylor expansion of $\Psi^{[1]}(\hat{x})$ around $z=0$.

Second, notice that thanks to the commutation relation \eqref{eq:SepComm}, the operator ${\Gamma}^\ast$ defined in \eqref{eq:Gamma_star_plus} anti-commutes with ${Q}_\mathrm{B}$. It can be also readily established that ${\Gamma}^*$ is nilpotent. We are therefore arriving at an algebra
\begin{subequations}
\label{eq:nilp_alg}
    \begin{align}
        \big({Q}_\mathrm{B}\big)^2 &=0\,,\\
        \big[{Q}_\mathrm{B},{\Gamma}^\ast\big]&=0\,,\\
        \big({\Gamma}^\ast\big)^2 &=0\,.
    \end{align}
\end{subequations}
We will find it convenient to denote
\begin{subequations}
\label{eq:Q_grades}
    \begin{align}
        Q^{[-1]} &\equiv -\Gamma^\ast\,,\\
         Q^{[0]} &\equiv +{Q}_\mathrm{B}\,,
    \end{align}
\end{subequations}
and set the remaining $Q^{[k]}$ for $k\neq 0,-1$ to zero. The algebra \eqref{eq:nilp_alg} can then be equivalently restated as
\begin{align}
    \sum_{l} Q^{[l]}Q^{[k-l]}=0
\end{align}
for $k=-2,-1,0$.

\subsection{Dimensional $\mathbb{Z}$-grading and multi-structures on the Hilbert space}

The integers appearing in the square brackets in superscripts in \eqref{eq:sf_repackaging} and \eqref{eq:Q_grades} will be seen to be associated with a $\mathbb{Z}$-graded structure. They will measure (up to a universal constant shift) the mass-dimension of the target fields. Since the fields $\Psi^{[k]}(\hat{x})$ all belong to the same worldsheet Hilbert space $\mathcal{H}$, their degree $k$ is directly related to the mass-dimension of the target fields $\Psi_s^{[k]}(x)$ which enter $\Psi^{[k]}(\hat{x})$ as expansion coefficients in an oscillator basis of $\mathcal{H}$. Beware that the mass-dimension degree $k$ of $\Psi^{[k]}(\hat{x})$ is completely orthogonal to its suspended Grassmann degree, which we keep denoting by $d(\Psi^{[k]})$.

\subsubsection{Multi-fields and multi-operators}

Having this in mind, we will find it convenient to organize string fields $\Phi^{[k]}$ having different mass-dimension degrees $k\in\mathbb{Z}$ into a multi-field
\begin{align}
    \mathbf{\Phi}(\hat{x}) = \sum_{k} \Phi^{[k]}(\hat{x})\,.\label{eq:multifield}
\end{align}
In the case of the dynamical string multi-field $\mathbf{\Psi}(\hat{x})$, only the degrees $k=1,2$ will be relevant for the action while the remaining $\Psi^{[k]}$ for $k\neq 1,2$ can be set to zero. Recall that the corresponding dynamical fields $\Psi^{[1]}$ and $\Psi^{[2]}$ were already constructed in \eqref{eq:sf_repackaging}.

Generic multi-fields $\mathbf{\Phi}$ can be acted upon by multi-operators $\mathbf{O}$ which again admit degree expansion
\begin{align}
    \mathbf{O} = \sum_k O^{[k]}\,.
\end{align}
This operator action can be naturally defined by writing
\begin{align}
    \mathbf{\Phi}\longrightarrow \mathbf{\Phi}' = \mathbf{O}\mathbf{\Phi}\,,
\end{align}
where the multi-field $\mathbf{\Phi}'$ can be degree-expanded in components $(\Phi')^{[k]}$ which satisfy
\begin{align}
    (\Phi')^{[k]} = \sum_l O^{[l]}\Phi^{[k-l]}\,.
\end{align}
In particular, given the notation \eqref{eq:Q_grades}, the operators ${Q}_\mathrm{B}$ and $\Gamma^\ast$ can be organized into the kinetic multi-operator
\begin{align}
    \mathbf{Q} = \sum_k Q^{[k]}\,,
\end{align}
where we set $Q^{[k]}=0$ for $k\neq 0,-1$.
The algebra \eqref{eq:nilp_alg} is then equivalent to the statement that $\mathbf{Q}$ is nilpotent, namely
\begin{align}
    \mathbf{Q}^2=0\,.
\end{align}
Also notice the degree of the components of $\mathbf{Q}$ indeed corresponds to their mass-dimension: while the combination $\alpha' \p^2$ appearing in $Q_\mathrm{B}$ has mass-dimension 0, the term $\alpha' \p_z$ appearing in $\Gamma_+^\ast$ has mass-dimension $-1$.

Furthermore, recognizing the gauge parameter $\Lambda(\hat{x})$ as a degree-$2$ field
\begin{align}
    \Lambda^{[2]}(\hat{x})\equiv \Lambda(\hat{x})
\end{align}
and organizing it into a multi-field $\mathbf{\Lambda}(\hat{x})$ where we set $\Lambda^{[k]}(\hat{x})$ for all $k\neq 0$, the gauge transformation of both the bulk string field $\Psi(\hat{x})$ and the boundary string fields $\Sigma_0(\hat{y})$, $\Sigma_1(\hat{y})$ can be encapsulated into writing
\begin{align}
    \delta_\mathbf{\Lambda}\mathbf{\Psi} = \mathbf{Q}\mathbf{\Lambda}\,.
\end{align}

\subsubsection{Symplectic form}

Since we want the string field theory action to have a definite mass-dimension (which we, conventionally, associate with degree 0), it follows that the symplectic form $\omega$ defined in Section \ref{sec:position} has to carry degree $-4$. That is, if we like, we can write $\omega^{[-4]}\equiv \omega$.
One can then define a symplectic form $\boldsymbol{\omega}_M$ on the space of multi-fields, which should be relevant for defining an SFT action, namely
\begin{align}   \boldsymbol{\omega}_M(\mathbf{\Phi}_1,\mathbf{\Phi}_2)\equiv \pi_0\,\omega\big(\mathbf{\Phi}_1,\boldsymbol{\theta}(\hat{x})\mathbf{\Phi}_2\big)\,.\label{eq:multiSF}
\end{align}
Here we have made use of the projector\footnote{In general, we can define projectors $\pi_k$ projecting on quantities with degree $k\in\mathbb{Z}$.} $\pi_0$ onto quantities of degree $0$. Furthermore, the components $\theta^{[k]}(\hat{x})$ of the multi-operator
\begin{align}
    \boldsymbol{\theta}(\hat{x}) = \sum_k \theta^{[k]}(\hat{x})
\end{align}
are defined by putting $\theta^{[k]}(\hat{x}_0)=0$ for $k<0$ and 
\begin{align}
    \theta^{[k]}(\hat{x}) = (-\p_z)^{k}\theta(\hat{x})\,,
\end{align}
where we recall that $\theta(x)$ is the indicator function pointing on the inside of $M$. Also note, that \eqref{eq:multiSF} indeed defines a bilinear product which is graded-antisymmetric with respect to the suspended Grassmann-grading, namely
\begin{align}
    \boldsymbol{\omega}_M(\mathbf{\Phi}_1,\mathbf{\Phi}_2)=-(-1)^{d(\mathbf{\Phi}_1)d(\mathbf{\Phi}_2)}\boldsymbol{\omega}_M(\mathbf{\Phi}_2,\mathbf{\Phi}_1)\,.
\end{align}
This is directly inherited from the corresponding property of $\omega$, given the fact that the operators $\theta^{[k]}(\hat{x})$ are BPZ-even for all $k$.

Let us now show that the kinetic multi-operator $\hat{\mathbf{Q}}$ is cyclic with respect to $\boldsymbol{\omega}_M$. First, notice that one can generalize \eqref{eq:Bdef} and \eqref{eq:Bdecomp} to write
\begin{align}
    [\theta^{[k]}(\hat{x}),\hat{Q}_\mathrm{B}] = \theta^{[k+1]}(\hat{x})\hat{\Gamma}^\ast +\hat{\Gamma}\, \theta^{[k+1]}(\hat{x})\,.
\end{align}
In parallel to the computation \eqref{eq:bpz_proof}, one can then show that the operator
\begin{align}
   \theta^{[k]}(\hat{x}) \hat{Q}_\mathrm{B} -\theta^{[k+1]}(\hat{x}) \hat{\Gamma}^\ast\label{eq:general_BPZ}
\end{align}
is BPZ odd. Finally, note that for any multi-fields $\mathbf{\Phi}_1$, $\mathbf{\Phi}_2$, we can expand
\begin{align}
    \boldsymbol{\omega}_M (\mathbf{\Phi}_1,\hat{\mathbf{Q}}\mathbf{\Phi}_2) = \sum_{\substack{i,j,k\\ i+j+k=4}} \omega\big(\Phi_1^{[i]},(\theta^{[k]}(\hat{x})\hat{Q}^{[0]}+\theta^{[k+1]}(\hat{x})\hat{Q}^{[-1]})\Phi_2^{[j]}\big)\,.
\end{align}
Recalling the identifications \eqref{eq:Q_grades}, it then immediately follows from the BPZ-oddness \eqref{eq:general_BPZ} that we can write
\begin{align}
    \boldsymbol{\omega}_M (\mathbf{\Phi}_1,\hat{\mathbf{Q}}\mathbf{\Phi}_2) = -(-1)^{d(\mathbf{\Phi}_1)}\boldsymbol{\omega}_M (\hat{\mathbf{Q}}\mathbf{\Phi}_1,\mathbf{\Phi}_2)\,,
\end{align}
thus concluding that the kinetic multi-operator $\hat{\mathbf{Q}}$ is indeed cyclic with respect to $\boldsymbol{\omega}_M$.

\subsection{Constrained action}

Given the definitions and properties outlined above, one can write down an action
\begin{align}
    S(\mathbf{\Psi}) = \frac{1}{2}\boldsymbol{\omega}_M(\mathbf{\Psi},\hat{\mathbf{Q}}\mathbf{\Psi})\,,\label{eq:unconstrainedS}
\end{align}
which, as a consequence of the nilpotency of the kinetic multi-operator $\mathbf{Q}$ and its cyclicity with respect to the symplectic form $\boldsymbol{\omega}_M$, is invariant under the gauge transformation $\delta_\mathbf{\Lambda}\mathbf{\Psi} = \hat{\mathbf{Q}}\mathbf{\Lambda}$. However, as the symplectic form $\boldsymbol{\omega}_M$ contains insertions of derivatives normal to $\p M$, it is not guaranteed that \eqref{eq:unconstrainedS} represents a well-defined variational principle. As we will confirm below, this is only achieved assuming that the dynamical string multi-field $\mathbf{\Psi}$ satisfies the gauge-invariant linear constraint
\begin{align}
 \theta^{[1]}(\hat{x}) \pi_0 \hat{\mathbf{Q}}\mathbf{\Psi}=0\,.\label{eq:lin_rel_multi}
\end{align}
At the same time, we will see that \eqref{eq:lin_rel_multi} ensures that the action \eqref{eq:unconstrainedS} is fully equivalent to the gauge-invariant action \eqref{eq:TheAction}.

\subsubsection{Equivalence with \eqref{eq:TheAction}}

First, let us note that the r.h.s.\ of \eqref{eq:unconstrainedS} can be expanded as
\begin{align}
    S(\mathbf{\Psi}) = S_{\mathrm{cyc},+}(\Psi)+\omega\big(\Psi,(\theta^{[1]}\hat{Q}_\mathrm{B}-\theta^{[2]}\hat{\Gamma}^\ast)\Psi^{[1]}\big)+\frac{1}{2}\omega\big(\Psi^{[1]},(\theta^{[2]}\hat{Q}_\mathrm{B}-\theta^{[3]}\hat{\Gamma}^\ast)\Psi^{[1]}\big)\,,\label{eq:S1}
\end{align}
where we have recalled the definition \eqref{eq:Scyc+} of the cyclic action $S_{\mathrm{cyc},+}(\Psi)$.
Substituting the $z$-expansion \eqref{eq:zexp} of the string field $\Psi^{[1]}$ and unfolding the $z$-derivatives from the insertions of $\theta^{[k]}$ onto the string fields using integration by parts on $\mathbb{R}^d$, one obtains, term by term,
\begin{subequations}
\label{eq:term_by_term}
\begin{align}
&\omega\big(\Psi,(\theta^{[1]}\hat{Q}_\mathrm{B}-\theta^{[2]}\hat{\Gamma}^\ast)\Psi^{[1]}\big)    =\nonumber\\
&\hspace{0.4cm}=\int_{\p M} d^{d-1} y \, \omega'\big(\Psi,\tilde{Q}\Sigma_0-\Omega_-^z\Sigma_1\big)+\int_{\p M}d^{d-1} y \, \omega'\big(\p_z\Psi,\alpha'c\Sigma_1+\Omega_+^z\Sigma_0\big)\,,\\
&\frac{1}{2}\omega\big(\Psi^{[1]},(\theta^{[2]}\hat{Q}_\mathrm{B}-\theta^{[3]}\hat{\Gamma}^\ast)\Psi^{[1]}\big)=\nonumber\\
&\hspace{0.4cm}=\int_{\p M} d^{d-1} y\,\omega'\big(\Sigma_1,\tilde{Q}\Sigma_0-\Omega_-^z\Sigma_0\big)+\int_{\p M} d^{d-1} y\,\omega'\big(\Sigma_2,\alpha'c\Sigma_1+\Omega_+^z\Sigma_0\big)\,.
\end{align}
\end{subequations}
At the same time, the constraint \eqref{eq:lin_rel_multi} is exactly equivalent to the linear relation \eqref{eq:lin_rel}, as we can write
\begin{align}
    -\Gamma^\ast \Psi^{[1]}(x)\big|_{z=0}=\alpha' c \Sigma_1(y)+\Sigma_0(y)\,.
\end{align}
Hence, we observe that as a consequence of the constraint \eqref{eq:lin_rel_multi}, the dangerous terms in \eqref{eq:term_by_term} containing $\p_z\Psi$ (as well as the one containing $\Sigma_2$), which would have spoiled the variational principle defined by \eqref{eq:unconstrainedS},  drop out. Substituting the results of \eqref{eq:term_by_term} back into \eqref{eq:S1}, we recover the action \eqref{eq:TheAction}.


\endgroup

\end{document}